\documentclass[preprintnumbers,amsmath,amssymb]{revtex4}
\usepackage{graphicx,subfigure}
\usepackage{dcolumn}
\usepackage{bm}
\usepackage{amssymb}
\usepackage{amsmath}

\usepackage[usenames, dvipsnames]{color}

\begin{document}

\title{The extended rigid body and the pendulum revisited}

\author{Manuel de la Cruz}
\email{fisikito@gmail.com}

\author{N\'{e}stor Gaspar}
\email{nex3t.gr@gmail.com}


\author{Rom\'{a}n Linares} 
\email{lirr@xanum.uam.mx}

\affiliation{Departamento de F\'{\i}sica, Universidad Aut\'onoma Metropolitana Iztapalapa,\\
San Rafael Atlixco 186, C.P. 09340, Ciudad de M\'{e}xico, M\'exico,
}

\date{\today}

\begin{abstract}
In this paper we revisit the construction by which the $SL(2,\mathbb{R})$ symmetry of the Euler equations allows to obtain the simple pendulum from the rigid body. We begin reviewing the original relation found by Holm and Marsden in which, starting from the two Casimir functions of the extended rigid body with Lie algebra $ISO(2)$ and introducing a proper momentum map, it is possible to obtain both the Hamiltonian and equations of motion of the pendulum. Important in this construction is the fact that both Casimirs have the geometry of an elliptic cylinder. By considering the whole  $SL(2,\mathbb{R})$ symmetry group, in this contribution we give all possible combinations of the Casimir functions and  the corresponding momentum maps that produce the simple pendulum, showing that this system can also appear when the geometry of one of the Casimirs is given by a hyperbolic cylinder and the another one by an elliptic cylinder. 
As a result we show that from the extended rigid body with Lie algebra $ISO(1,1)$, it is possible to obtain the 
pendulum but only in circulating movement. Finally, as a by product of our analysis we provide the momentum maps 
that give origin to the pendulum with an imaginary time. Our discussion covers both the algebraic and the geometric 
point of view.
\end{abstract}

\maketitle

\section{Introduction}

The simple pendulum and the torque free rigid body are two well understood physical systems in both classical and 
quantum mechanics. The first systematic study of the pendulum is attributed to Galileo Galilei around 1602 and its 
dynamical description culminated  with the development of the elliptic functions by Abel \cite{Abel} and Jacobi 
\cite{Jacobi,JacobiFundamenta}, which turn out to be the analytical solutions to the equation of motion of the pendulum 
(for a review of elliptic functions see for instance \cite{Whittaker:1917,DuVal,Lang,Lawden,McKean,Armitage}  and 
\cite{Belendez,Ochs,Linares} for the solutions of the pendulum). The quantization of the pendulum is based on the 
equivalence  between the Schr\"{o}dinger equation and the Mathieu differential equation, result developed originally by 
Condon in 1928 \cite{PhysRev.31.891} and source of subsequent  analysis of different aspects of the quantum system 
\cite{doi:10.1119/1.1987121,doi:10.1119/1.12332}. On the other hand in 1758 Euler showed that the equations of 
motion that describe the rotation of a rigid body form a vectorial quasilinear first-order ordinary differential equations set 
\cite{Euler}. A geometric construction of the solution was given later on by Poinsot \cite{Poinsot} and analytically these 
solutions are given, as for the simple pendulum, by elliptic functions (see for instance 
\cite{Landau,MarsdenRatiu,Pinna,Holm} and references therein). The quantization of the problem was attacked first by 
Kramers and Ittmann \cite{Ittmann} and since then many authors have contributed to understand deeper many aspects 
of the problem \cite{King,spence,LukacSmoro,Patera,Pina1999159,Pina,LeyKoo,MendezFragoso2015115}.

Despite the old age of these problems, from time to time there are some new physical aspects uncovered about these 
systems that contribute to our knowledge and understanding of physics in general. The list is long and here we point 
out just four examples: i) In 1973 Y. Nambu, taking the Liouville theorem as a guiding  principle, proposed a 
generalization of the classical Hamiltonian dynamics by supersede the usual two dimensional phase space 
with a $n$-dimensional one  \cite{Nambu:1973qe}. The dynamics in the new phase space was formulated via an 
$n$-linear fully antisymmetric Nambu bracket with two or more ``Hamiltonians"; as an example, Nambu applied his 
formalism to the free rigid body. ii) Hinted by an article by Deprit \cite{doi:10.1119/1.1974113} and using the 
$SL(2,\mathbb{R})$ symmetry of the Euler equations, in 1991 Holm and Marsden built a new Hamiltonian in such a 
way that the new dynamical system, so called in the literature {\it extended rigid body},  can be written as a Lie-Poisson 
system whose different Lie algebras structures can be $SO(3)$, $ISO(2)$ or Heis$_3$ \cite{Marsden1}; particularly  
interesting is the $ISO(2)$ case, where the phase space of the Eulerian top is filled with invariant elliptic cylinders, on 
each of which, the dynamics, in elliptic  coordinates, is the dynamics of a standard simple pendulum. iii) In 1995 R. 
Montgomery  computed the change in the geometric phase for the attitude of the rigid body when the angular 
momentum vector in the body frame performs one period of its motion \cite{Montgomery}. iv) Finally in 2017 
\cite{VanDamme}  the free rotation of a classical rigid body was used in the control of two-level quantum systems by 
means of external electromagnetic pulses. In particular, authors showed that the dynamics of a rigid body can be used 
to implement one-qubit quantum gates. 

In this paper we are interested in explore deeper the relation between the extended rigid body and the simple 
pendulum.  As we have mentioned above, in the original paper \cite{Marsden1} Holm and Marsden showed that the 
different Lie algebras structures of the extended rigid body are $SO(3)$, $ISO(2)$ or Heis$_3$, however in 
\cite{Iwai2010501} authors showed that the complete list of possible Lie algebras are all the ones related to $SO(3)$ 
via analytical continuation and group contractions, which means that the algebra $ISO(1,1)$ must be also included. 
Even more, according to \cite{Marsden1} the pendulum can be obtained from the extended rigid body if, from the 
geometrical point of view, the surfaces of the two new Casimir functions have the shape of elliptic cylinders, which 
leads to the fact that the corresponding Lie algebra is $ISO(2)$. Very recently using a representation of the rigid body 
in terms of two free parameters $e_0$ and $\kappa$ \cite{delaCruz:2017qkh} instead of the usual five parameters: 
energy $E$, magnitude of the angular momentum $L$, and the three principal moments of inertia $I_1$, $I_2$ and 
$I_3$, the classification of the inequivalent $SL(2,\mathbb{R})$ combinations of the Casimir functions  was discussed 
both algebraically and geometrically. It turns out that whereas the geometry of the Casimir function that 
represents the square of the angular momentum continues being an $S^2$ sphere, the geometry of the Casimir 
function associated to the kinetic energy which is an ellipsoid in the five parameters representation of the rigid body, is 
replaced by an elliptic hyperboloid that can be either of one or two sheets depending on the numerical values of both 
$e_0$ and $\kappa$, making the classification of the $SL(2,\mathbb{R})$ combinations of the Casimir functions richer. 
A result of this paper is to show that, considering all possible different geometries of the new Casimir functions, it is 
possible to obtain the pendulum also when the geometries associated to them are an elliptic cylinder and a hyperbolic 
cylinder. Specifically, when the cotangent space of the simple pendulum is given by the elliptic cylinder, the Lie algebra 
of the Hamiltonian vector fields associated to the coordinates in the rigid body-fixed frame is $ISO(2)$, whereas if the 
cotangent space is given by the hyperbolic cylinder, the Lie algebra is $ISO(1,1)$. Even more, we show 
explicitly that for the $ISO(2)$ case we always obtain the whole set of motions of the pendulum whereas for the  
$ISO(1,1)$ case we get the circulating motions but not the oscillatory ones. To the best of our knowledge this 
result has not been discussed previously in the literature. 
 
Our exposition is as self-contained as possible. Section \ref{RigidBody} is dedicated to summarize the main 
characteristics of the rigid body and its equations of motion, specially the $SL(2,\mathbb{R})$ symmetry 
of the later. In section \ref{pendulum} we discuss the Hamiltonian structure of the simple pendulum as function of 
both a real time and an imaginary time.  The study of the whole $SL(2,\mathbb{R})$ transformations can be 
divided in three general sets where each set contains different fixed forms of  the $SL(2,\mathbb{R})$ matrices.  
One of the three sets does not give origin to the pendulum and therefore we focus on the other two. Section 
\ref{Buena} is devoted to the set of matrices where the relation between the rigid body and the simple pendulum can 
be stablished. This set can be subdivided in two general cases determined by the geometry of the Casimir functions, 
subsection  \ref{TwoElliptic} is dedicated to the case where both Casimir functions are given geometrically by elliptic 
cylinders whereas subsections \ref{OscPenIntHypEllip} and \ref{SecDeg} are dedicated to the study of the cases where 
one Casimir function has the geometry of an elliptic cylinder and the another Casimir function has the geometry of a 
hyperbolic cylinder. In section \ref{CNeq0D0} we discuss the third and last set; as we will argue, this set is a limiting 
case of the one in section \ref{Buena} and therefore there does also exist a relation between the simple pendulum and 
the rigid body. Our conclusions are given in section \ref{conclusions}.


\section{The Euler equations and its $SL(2,\mathbb{R})$ symmetry} \label{RigidBody}


The torque free rigid body motion is one of the best understood systems in physics and the amount of papers and 
books discussing their dynamical properties is overwhelming. However there are still some properties related to this  
system that deserve to be explored further. In this section we give a short summary of the characteristics of the system 
that are relevant for our analysis of the relation between the so called {\it extended rigid body} and  the 
{\it simple pendulum}. We base our discussion in \cite{Pinna,Marsden1,MarsdenRatiu,delaCruz:2017qkh}.

\subsection{The Euler equations and the Casimir functions}

It is well known that in a body-fixed reference frame the motion of the rigid body is governed by the Euler equations 
(see for instance \cite{MarsdenRatiu})
\begin{equation}\label{EulerL}
\frac{d \vec{L}}{dt} = \vec{L} \times I^{-1}  \vec{L},
\end{equation}
where $\vec{L}$ is the vector of angular momentum and $I$ is the moment of inertia tensor. If  the body-fixed frame is 
oriented to coincide with the principal axes of inertia, the tensor $I$ is diagonal. Without losing generality in the 
following discussion we consider that the principal moments of inertia satisfy the inequality
\begin{equation}\label{OrderI}
I_1 < I_2 < I_3.
\end{equation}
When written in a basis, the $L_i$  components of the angular momentum are the generators of a  $SO(3)$ Lie 
algebra
\begin{equation}\label{LieAlgebra}
[L_i,L_j]=\varepsilon_{ijk} L_k.
\end{equation}
The system has two Casimir functions, the rotational kinetic energy $C_1$ and the square of the angular momentum 
$C_2$, which are given in terms of the moments of inertia and the components of $\vec{L}$ as
\begin{equation}
C_1(L_1, L_2, L_3) \equiv \frac{L_1^2}{2I_1}+  \frac{L_2^2}{2I_2}+ \frac{L_3^2}{2I_3}, \label{ConE}
\end{equation}
and
\begin{equation}
C_2(L_1,L_2,L_3) \equiv  L_1^2+L_2^2+L_3^2,\label{ConL}
\end{equation}
respectively. The dynamical  problem is usually solved using the Poinsot construction, in an Euclidean space 
$\mathbb{R}^3$ whose coordinates are the components of the angular momentum. When the vector $\vec{L}$ moves 
relative to the axes of inertia of the top, it  lies along the curve of intersection of the surfaces $C_1=E$ = constant (an 
ellipsoid with semiaxes $\sqrt{2EI_1}$, $\sqrt{2EI_2}$ and $\sqrt{2EI_3}$) and $C_2=L^2$ = constant (a sphere of 
radius $L$). Because $2EI_1< L^2 < 2EI_3$, the radius of the sphere has a value between the minimum and 
maximum values of the semiaxes of the ellipsoid. At first sight the solutions of the Euler equations (\ref{EulerL}) depend 
on five different parameters, the three moments of inertia, the energy $E$ and the square of the angular momentum 
$L^2$. However it has been shown that the problem can be rewritten in such a way that the solutions depend   
only on two parameters \cite{Ittmann,LukacSmoro,Pinna,Pina1999159}. The first parameter $e_0$ is related to the 
quotient $E/L^2$ whereas the second parameter $\kappa$ codifies the values of the three moments of inertia, 
specifically, if ${\cal I}$ is the diagonal matrix ${\cal I}$= diag$(1/I_1, 1/I_2, 1/I_3)$, then the relation between the five 
parameters: $\{ I_i\}$, $E$ and $L^2$ with the four parameters: $\{ e_i\}$ and $e_0$ is given by
\begin{equation}
\frac{1}{I_i}-\frac13 \mbox{Tr}  \, {\cal I} \equiv \sqrt{\frac{g_2}{3}} e_i \hspace{0.8cm} 
\frac{2E}{L^2} - \frac13  \mbox{Tr} \, {\cal I} \equiv \sqrt{\frac{g_2}{3}} e_0,
\end{equation}  
with $i=1,2,3$ and
\begin{equation}\label{DefA}
g_2 \equiv  \frac{2}{3} \left[  \left( \frac{1}{I_1}- \frac{1}{I_2} \right)^2 + \left( \frac{1}{I_2}- \frac{1}{I_3} \right)^2   +
\left( \frac{1}{I_3}- \frac{1}{I_1} \right)^2  \right] >0 .
\end{equation}
Notice that the new four parameters are dimensionless, and abusing of the language we can call to the set $\{e_ i\}$ 
{\it dimensionless inertia parameters}, analogously although $e_0$ is not the energy of the system we can 
call it the {\it dimensionless energy parameter}.  The dimensionless inertia parameters can be written in terms of only 
one angular parameter $\kappa$ in the form
\begin{equation}\label{eidek}
e_1= \cos(\kappa) \,, \qquad e_2= \cos(\kappa-2\pi/3)\,,  \qquad  e_3= \cos(\kappa+2\pi/3), 
\end{equation}
and they are restricted to satisfy the conditions
\begin{equation}\label{CondDimless}
e_1+e_2+e_3=0, \hspace{0.5cm} e_1^2+e_2^2+e_3^2=3/2 \hspace{0.3cm} \mbox{and} \hspace{0.3cm} 
e_1e_2e_3=(\cos 3 \kappa) /4 .
\end{equation}
Geometrically, in the three dimensional space $\{ e_1,e_2,e_3\}$, the first condition in (\ref{CondDimless}) represents a 
plane crossing the origin and the second condition represents a sphere of radius $\sqrt{3/2}$. The intersection of these 
two surfaces is a circle which is parameterized by the angular parameter $\kappa$. The role of $g_2$ (equation 
(\ref{DefA})) is to change the size of the sphere and therefore the size of the intersecting circle. It is clear that given a 
specific rigid body or equivalently a set of values for the three principal moments of inertia $I_i$, the value of the 
angular parameter $\kappa$ is completely determined.

In general $\kappa \in [0,2\pi]$, but notice that the condition (\ref{OrderI}) in terms of the dimensionless inertia 
parameters is obtained if the free parameter $\kappa$ takes values in the subinterval $\kappa \in (0,\pi/3)$, i.e. 
\begin{equation}\label{esOrdering}
e_3 < e_2 < e_1.
\end{equation}
The other five subintervals of length $\pi/3$ produce the other five possible orders for the $e_i$'s, for instance, if  
$\kappa \in (\pi/3,2\pi/3)$ the parameters ordering is $e_3<e_1<e_2$, etc. For any value of $\kappa$, at least one of 
the inertia parameters is positive, another is negative and the third one may be either, positive, null or negative. 
Throughout  all the paper we work in the interval $\kappa \in (0,\pi/3)$, for which $e_1>0$, $e_3<0$ and $e_2$ 
can be either positive, negative or null.
 
Finally introducing the dimensionless coordinates $u_i \equiv L_i/L$, the Euler equations (\ref{EulerL}) are rewritten as
\begin{equation}
\frac{d u_i}{dt}  = \frac12  \sqrt{\frac{g_2}{3}}  L \, \varepsilon_{ijk} (e_k-e_j) u_j u_k .
\end{equation}
It is possible to absorb the factor $ \sqrt{\frac{g_2}{3}}  L$, by defining a dimensionless time parameter: 
$x \equiv t  \sqrt{\frac{g_2}{3}}  L$, obtaining 
\begin{equation}\label{euleradim}
\dot{u}_i =\frac12  \varepsilon_{ijk} (e_k-e_j) u_j u_k \,,
\end{equation}
where ($\cdot$) denotes derivative with respect to the dimensionless time $x$. In a similar fashion Casimir functions 
(\ref{ConE}) and (\ref{ConL}) read now as
\begin{eqnarray}
C_1(u_1,u_2,u_3) & \equiv &  e_1 u_1^2 + e_2 u_2^2 + e_3u_3^2,                                                     \label{Con2L} \\
C_2(u_1,u_2,u_3) & \equiv & u_1^2+u_2^2+u_3^2.
\label{Con2E}
\end{eqnarray}
In the dimensionless angular momentum space $\mathbb{R}^3$: ($u_1,u_2,u_3$), $C_2=1$ and it 
represents a unitary sphere, whereas for the other Casimir: $C_1=e_0$ and the ellipsoid (\ref{ConE}), is replaced by an elliptic hyperboloid that can be either of one or two sheets depending on the numerical values of both $2E/L^2$ and $\kappa$, or equivalently on the number of positive and negative coefficients (dimensionless inertia parameters) in the equation. In this latter case the Casimir surface can also have the geometry of an elliptic cone or a hyperbolic cylinder in the proper limit situations. A complete classification of the geometrical shapes of the Casimir surface $C_1$ can be 
found in \cite{delaCruz:2017qkh}. 

Regarding the solutions of the Euler equations (\ref{euleradim}), given the ordering (\ref{esOrdering}) of the 
dimensionless inertia parameters, the solutions depend of the relative value between $e_0$ and $e_2$. Here 
we are not going into details, but for purposes of completeness in our exposition, we present the explicit solutions, 
details can be found for instance in \cite{Pinna}.
\begin{itemize}
\item[{\bf I.}] Case $e_3<e_2<e_0<e_1 $ \, ({\it i.e.} $1/I_2<2E/L^2$).
\end{itemize}
When the energy parameter $e_0$ is between the inertia parameters $e_2$ and $e_1$, the solutions are given by 
\begin{equation}\label{Solutions1}
u_1(\tau)=  \mbox{sn}(\tau ' ,m_c) \, \mbox{dn}(\tau,m), \hspace{0.4cm}
u_2(\tau)= \mbox{dn}(\tau ' ,m_c)  \, \mbox{sn}(\tau,m), \hspace{0.4cm}
u_3 (\tau)= \mbox{cn}(\tau ' ,m_c) \,  \mbox{cn}(\tau,m), 
\end{equation}
where $\tau$ is a dimensionless time parameter defined as
\begin{equation}\label{Deftau}
\tau= x \sqrt{(e_1-e_2)(e_0-e_3)}.
\end{equation}
Here the amplitudes of the solutions are written as Jacobi elliptic functions at parameter $\tau ' =$ constant and are 
related to the dimensionless parameters $e_i$ and $e_0$ in the form
\begin{equation}\label{identifications}
\mbox{sn}^2(\tau ' ,m_c) = \frac{e_0-e_3}{e_1-e_3} , \hspace{0.5cm}
\mbox{cn}^2(\tau ' ,m_c) = \frac{e_1-e_0}{e_1-e_3} , \hspace{0.5cm}
\mbox{dn}^2(\tau ' ,m_c)  = \frac{e_1-e_0}{e_1-e_2}.
\end{equation}
The square modulus $m$ and the complementary modulus $m_c$ in the equation (\ref{Solutions1}) are given by
\begin{equation}\label{equ}
m^2\equiv \frac{(e_2-e_3)}{(e_0-e_3)}  \frac{(e_1-e_0)}{(e_1-e_2)}  = \frac{(e_1-e_0)}{(e_0-e_3)}\, 
\frac{k_1^2}{k_2^2}, \hspace{0.3cm} \mbox{and}  \hspace{0.3cm} 
m_c^2 \equiv \frac{(e_0-e_2)}{(e_0-e_3)}  \frac{(e_1-e_3)}{(e_1-e_2)}= \frac{(e_0-e_2)}{(e_0-e_3)}\, \frac{1}{k_2^2},
\end{equation}
which take values in the interval $0<m^2<1$ and $0<m_c^2<1$ and satisfy $m^2+m_c^2=1$. Here the quotients
\begin{equation}\label{defke}
k_1^2=\frac{e_2-e_3}{e_1-e_3} ,
 \hspace{0.5cm} \mbox{and}  \hspace{0.5cm}
k_2^2=\frac{e_1-e_2}{e_1-e_3},
\end{equation}
satisfy in a similar way $k_1^2+k_2^2=1$.
\begin{itemize}
\item[{\bf II.}] Case $e_3<e_0<e_2<e_1$ \, ({\it i.e.}  $ 2E/L^2<1/I_2$).
\end{itemize}
In this case the solutions are given by 
\begin{equation}\label{Solutions2}
u_1=\mbox{cn}\left( \tau ' , i\frac{m_c}{m} \right) \,  \mbox{cn}\left( m\tau,\frac{1}{m} \right),\hspace{0.3cm}
u_2=\mbox{dn} \left( \tau ' , i\frac{m_c}{m} \right)  \, \mbox{sn} \left( m\tau,\frac{1}{m} \right), \hspace{0.3cm}
u_3 =\mbox{sn}\left( \tau ' , i\frac{m_c}{m} \right) \, \mbox{dn}\left( m\tau,\frac{1}{m} \right),
\end{equation}
where the square modulus of the elliptic function take values in the interval
$0<1/m^2 < 1$, with $m^2$ as defined in (\ref{equ}) but due to the fact that $e_3<e_0<e_2<e_1$, 
$m^2>1$. The time parameter  $\tau$ is the one defined in (\ref{Deftau}), whereas the amplitudes are written as Jacobi 
elliptic functions at parameter $\tau ' =$ constant
\begin{equation}\label{identifications2}
\mbox{sn}^2\left( \tau ' , i\frac{m_c}{m} \right) = \frac{e_1-e_0}{e_1-e_3} , \hspace{0.5cm}
\mbox{cn}^2 \left( \tau ' , i\frac{m_c}{m} \right) = \frac{e_0-e_3}{e_1-e_3} , \hspace{0.5cm}
\mbox{dn}^2 \left( \tau ' , i\frac{m_c}{m} \right)  = \frac{e_0-e_3}{e_2-e_3}.
\end{equation}
Physical interpretation of the solutions is straightforward, the curves (\ref{Solutions1}) and (\ref{Solutions2}) are the 
parameterization of the intersection of the unitary sphere of angular momentum and the corresponding surface 
of the Casimir function (\ref{Con2L}) \cite{delaCruz:2017qkh} (see figure \ref{GeometryFig}). 

At this point it is convenient to make clear the possible values that can take the square modulus $m^2$, the 
complementary modulus $m_c^2$, their inverse values and the quotients of these quantities. The reason for it, is 
that these quantities will appear when the connection between the rigid body and the simple pendulum be established.
\begin{table}[htb]  
\begin{tabular}{|c | c | c | c | c | c | c |} 
\hline
 & $m^2$ & $m_c^2$ & $\frac{1}{m^2}$ & $-\frac{m_c^2}{m^2}$ & $\frac{1}{m_c^2}$& $-\frac{m^2}{m_c^2} $\\
\hline
\hline
$e_2 < e_0$ & $(0,1)$       &  $(0,1)$       & $(1,\infty)$ & $(-\infty,0)$ &  $(1,\infty)$ & $(-\infty,0)$ \\
$e_0 < e_2$ & $(1,\infty)$ & $(-\infty,0)$  & $(0,1)$      & $(0,1)$        &  $(-\infty,0)$& $(1,\infty)$  \\
 \hline
\end{tabular}
\caption{Intervals of values of the square modulus $m^2$, complementary modulus $m_c^2$, their inverse values $1/m^2$, $1/m_c^2$ and the quotients $m^2/m_c^2$ and $m_c^2/m^2$, as functions of the relative values between the dimensionless energy parameter $e_0$ and the dimensionless inertia parameter $e_2$.}\label{valuesSM}
\end{table}

Notice that, in an analogous way to the fact that $m^2$ and $m_c^2$ are complementary to each other in the sense that $m^2+m_c^2=1$, the couple $1/m^2$ and $-m_c^2/m^2$ are complementary to each other, as well as the couple 
$1/m_c^2$ and $-m^2/m_c^2$, i.e.
\begin{equation}\label{complementary}
m^2+m_c^2=1, \hspace{1cm} \frac{1}{m^2}-\frac{m_c^2}{m^2}=1, \hspace{1cm}  \frac{1}{m_c^2}-\frac{m^2}{m_c^2}=1.
\end{equation}

\subsection{$SL(2 ,{\rm I\!R})$ symmetries of the Euler equations}\label{Gauge}

In order to make manifest the gauge symmetries of the Euler equations, we notice that its dimensionless  form
\begin{equation}
\dot{\vec{u}}= \vec{u} \times \epsilon \vec{u}
\end{equation}
with $\epsilon$ a diagonal matrix of the form $\epsilon =$diag$(e_1,e_2,e_3)$, can be rewritten as the gradient of 
two scalar functions
\begin{equation}
\dot{\vec{u}}=\nabla l  \times \nabla h,
\end{equation}
where
\begin{eqnarray}
h\equiv \frac12 C_1 (u_1, u_2, u_3)- \frac12 e_0 =0 &\Rightarrow&  \epsilon \vec{u}= \nabla h = 
 (e_1u_1,e_2u_2,e_3u_3), \, \, \, \, \, \, \label{Defh}\\
l \equiv \frac12 C_2 (u_1, u_2, u_3) - \frac12 =0 &\Rightarrow& \vec{u}=\nabla l = (u_1,u_2,u_3) \label{Defl}.
\end{eqnarray}
This form of writing the Euler equations makes explicit its invariance under any $SL(2, {\rm I\!R})$  transformation 
\cite{Marsden1}.
It is straightforward to check that the transformation
\begin{equation}\label{TransSL2}
 \left( 
\begin{array}{c}
{\cal H} \\
{\cal N}
\end{array}
\right) = \left( 
\begin{array}{cc}
a & b \\
c& d
\end{array}
\right)
 \left( 
\begin{array}{c}
h \\
l
\end{array}
\right), 
\end{equation}
with $ad-bc=1$ leads to
\begin{equation}\label{EulerHN}
\dot{\vec{u}}=\nabla {\cal N}  \times \nabla {\cal H}.
\end{equation}
At this point we can consider either of the new Casimir surfaces  ${\cal N}$  or ${\cal H}$ as the Hamiltonian 
surface. The systems that have this property are called bi-hamiltonian (see for instance 
\cite{Nambu:1973qe,MarsdenRatiu}). Usually the one that is chosen as the hamiltonian surface is ${\cal H}$, 
thus, given a dynamical system with Hamiltonian $H$, any dynamical quantity $Q$ evolves with time according to
\begin{equation}
\dot Q=X_{\cal H}Q,
\end{equation}
where the generator $X_G$ is given by
\begin{equation}
X_G=( \nabla {\cal N}  \times \nabla G) \cdot \nabla.
\end{equation}
Notice that we are denoting with a different letter to the Hamiltonian or Casimir function: 
$H=H(u_1,u_2,u_3)$ and to the Casimir surface: ${\cal H} (u_1,u_2,u_3) \equiv H (u_1,u_2,u_3)- E=0$.

According to the transformation (\ref{TransSL2}), the generator $X_G$ in components form is expressed as
\begin{eqnarray}
X_G &=&(ce_1+d)u_1\left( \frac{\partial G}{\partial u_2}\frac{\partial}{ \partial u_3} -  
\frac{\partial G}{\partial u_3}\frac{\partial}{ \partial u_2} \right) + 
(ce_2+d)u_2\left( \frac{\partial G}{\partial u_3}\frac{\partial}{ \partial u_1} -  
\frac{\partial G}{\partial u_1}\frac{\partial}{ \partial u_3} \right) \nonumber \\
& & + (ce_3+d)u_3\left( \frac{\partial G}{\partial u_1}\frac{\partial}{ \partial u_2} -  
\frac{\partial G}{\partial u_2}\frac{\partial}{ \partial u_1} \right).
\end{eqnarray}
In order to determine the structure of the Lie algebra, we calculate the Lie-Poisson brackets for the Hamiltonian vector fields associated with the coordinate functions $u_i$
\begin{eqnarray}
X_{u_1}&=& (ce_3+d) u_3 \partial_2- (ce_2+d) u_2 \partial_3, \nonumber \\
X_{u_2}&=& (ce_1+d) u_1 \partial_3- (ce_3+d) u_3 \partial_1, \label{BasicGen} \\
X_{u_3}&=& (ce_2+d) u_2 \partial_1- (ce_1+d) u_1 \partial_2. \nonumber
\end{eqnarray}
Explicitly, the Lie-Poisson brackets  are given by
\begin{equation}\label{GeneralLie}
[X_{u_1},X_{u_2}]=(ce_3+d)X_{u_3}, \hspace{0.1cm} 
[X_{u_2},X_{u_3}]=(ce_1+d)X_{u_1}, \hspace{0.1cm}
[X_{u_3},X_{u_1}]=(ce_2+d)X_{u_2}.
\end{equation}
Regarding the classification of the different Lie algebras that arise from the $SL(2, {\rm I\!R})$ invariant orbits, it turns out that, besides the $SO(3)$ Lie algebra (\ref{LieAlgebra}) which corresponds to the rigid body, the algebras 
$SO(2,1)$, $ISO(2)$, $ISO(1,1)$ and Heis$_3$, also emerge. The systems corresponding to these algebras are termed under the name {\it extended rigid bodies}. The whole classification can be found, for instance, in 
\cite{delaCruz:2017qkh}.

Because is relevant for our discussion, it is important to emphasize that if instead we take the Casimir surface 
${\cal N}$ as the Hamiltonian surface, then the generator $X_G$ is given by
\begin{equation}
X_G= - ( \nabla {\cal H}  \times \nabla G) \cdot \nabla \hspace{0.5cm} \Rightarrow \hspace{0.5cm} \dot Q=X_{\cal N}Q.
\end{equation}
Clearly $X_{\cal N}=-X_{\cal H}$. If we denote as $\tilde{X}_{u_i}$ to the Hamiltonian vector fields associated to the 
coordinates, we obtain in this case
\begin{eqnarray}
\tilde{X}_{u_1}&=& -(ae_3+b) u_3 \partial_2+ (ae_2+b) u_2 \partial_3, \nonumber \\
\tilde{X}_{u_2}&=& -(ae_1+b) u_1 \partial_3+ (ae_3+b) u_3 \partial_1, \label{BasicGenN} \\
\tilde{X}_{u_3}&=& -(ae_2+b) u_2 \partial_1+ (ae_1+b) u_1 \partial_2. \nonumber
\end{eqnarray}
With these vector fields the Lie-Poisson brackets are given explicitly by
\begin{equation}\label{GeneralLieN}
[\tilde{X}_{u_1},\tilde{X}_{u_2}]=-(ae_3+b)\tilde{X}_{u_3}, \hspace{0.1cm} 
[\tilde{X}_{u_2},\tilde{X}_{u_3}]=-(ae_1+b)\tilde{X}_{u_1}, \hspace{0.1cm}
[\tilde{X}_{u_3},\tilde{X}_{u_1}]=-(ae_2+b)\tilde{X}_{u_2}.
\end{equation}
The classification of the different Lie-Algebras coincide with the five mentioned above. Details are completely 
analogous to the previous ones and we do not discuss them further.

\section{The pendulum}\label{pendulum}

As in the case of the extended rigid body, solutions of the simple pendulum system are given in terms of Jacobi elliptic 
functions which are defined in the whole complex plane $\mathbb{C}$ and are doubly periodic. Due to the relevance 
of the expressions of the pendulum energy in both real and imaginary time, for the purposes of this paper, 
in this section we review briefly the main characteristics of the mathematical formulation of the simple pendulum. This 
is a well understood system and there are many interesting papers and books on the subject 
\cite{Whittaker:1917,Lawden,Armitage,Belendez, Ochs,Brizard,Appell2,Helmholtz}. In our discussion we follow 
\cite{Linares} mainly.

\subsection{Real time pendulum and solutions}

Let us start considering a pendulum of point mass $m$ and length $r$, in a constant downwards gravitational field, of 
magnitude $-g$ ($g>0$). If $\theta$ is the polar angle measured counterclockwise respect to the vertical line and 
$\dot{\theta}$ stands for the time derivative of this angular position, the lagrangian of the system is given by
\begin{equation}
L(\theta,\dot{\theta})= \frac{1}{2}mr^2 \dot{\theta}^2-mgr(1-\cos \theta).
\end{equation}
Here the zero of the potential energy is set at the lowest vertical position of the pendulum ($\theta =2 n \pi$, 
with $n \in \mathbb{Z}$). The equation of motion for this system is 
\begin{equation}\label{eqmotion}
\ddot{\theta}+\frac{g}{r} \sin \theta = 0,
\end{equation}
which after integration gives origin to the conservation of total mechanical energy
\begin{equation}\label{EnergyPen}
E=\frac12 m r^2 \dot{\theta}^2+2mgr\sin^2 \left( \frac{\theta}{2} \right)=constant.
\end{equation}
Physical solutions of equation  (\ref{EnergyPen}) exist only if $E \geq 0$. We can rewrite the equation in dimensionless 
form, in terms of the dimensionless energy parameter: $k_E^2 \equiv \frac{E}{2mgr}$, and the dimensionless real time 
variable: 
$x \equiv \sqrt{\frac{g}{r}} t \in \mathbb{R}$, obtaining 
\begin{equation}\label{DimensionlessE}
\left( \frac{p}{2} \right)^2 + \sin^2 \left( \frac{\theta}{2}\right) = k_E^2,
\end{equation}
where $p$ is the dimensionless angular velocity: $p(x) \equiv d\theta/ d x$. By inspection of the potential we conclude 
that the pendulum has four different types of solutions depending of the value of the constant $k_E^2$:\\

$\bullet$ {\it Static equilibrium} ($\dot{\theta}=0$): The trivial behavior occurs when either $k_E^2=0$ or 
$k_E^2=1$. In the first case, necessarily $\dot{\theta}=0$. For the case $k_E^2=1$ we consider also the 
situation where $\dot{\theta}=0$. In both cases, the pendulum does not move, it is in static equilibrium. When 
$\theta =2 n \pi$ the equilibrium is stable and when  $\theta = (2n+1) \pi$ the equilibrium is unstable. \\

$\bullet$ {\it  Oscillatory motions} ($ 0< k_E^2<1$): In these cases the pendulum swings to and fro, 
respect to a point of stable equilibrium. The analytical solutions are given by
\begin{eqnarray}
\theta (x)&=&2  \arcsin [k_E \,  \mbox{sn}(x-x_0,k_E)], \label{OscSol} \\
p(x) &=& 2 \, k_E \,  \mbox{cn}(x-x_0,k_E), \label{OscP}
\end{eqnarray} 
where the square modulus $k^2$ of the elliptic functions is given directly by the energy parameter: $k^2 \equiv k^2_E$. 
Here $x_0$ is a second constant of integration and appears when equation (\ref{DimensionlessE}) is integrated out. It 
means physically that we can choose the zero of time arbitrarily. The period of the movement is $4K$, or restoring the 
dimension of time, $4K\sqrt{g/r}$, with $K$ the quarter period of the elliptic function sn$(x-x_0,k_E)$. \\

$\bullet$ {\it Asymptotical motion} ($k_E^2=1$ and $\dot{\theta}\neq0$): In this case the angle $\theta$ 
takes values in the open interval  $(-\pi, \pi)$ and therefore, $\sin(\theta/2)  \in (-1,1)$. The particle just 
reach the highest point of the circle. The analytical solutions are given by
\begin{eqnarray}
\theta (x)&=& \pm  2  \arcsin [\tanh(x-x_0)], \label{AsinSol} \\
p(x) &=& \pm 2 \, \mbox{sech}(x-x_0).
\end{eqnarray} 
The sign $\pm$ corresponds to the movement from $(\mp \pi \rightarrow \pm \pi)$.  Notice that $\tanh(x-x_0)$, 
takes values in the open interval $ (-1,1)$ if: 
$x-x_0 \in (-\infty, \infty)$. For instance if $\theta \rightarrow \pi$, $x-x_0 \rightarrow \infty$ 
and $ \tanh(x-x_0)$ goes asymptotically to 1. It is clear that this movement is not periodic. In the 
literature it is common to take $x_0=0$.\\

$\bullet$ {\it Circulating motions} ($k_E^2>1$): In these cases the momentum of the particle is large 
enough to carry it over the highest point of the circle, so that it moves round and round the circle, always in 
the same direction. The solutions that describe these motions are of the form
\begin{eqnarray}
\theta (x)&=& \pm 2\,  \mbox{sgn}\left[\mbox{cn} \left( k_E (x-x_0) , \frac{1}{k_E} \right) \right] 
 \arcsin \left[ \mbox{sn} \left( k_E (x-x_0) , \frac{1}{k_E} \right) \right], 
\label{CirSol} \\
 p(x) &=& \pm 2 \, k_E \, \mbox{dn} \left( k_E (x-x_0) , \frac{1}{k_E} \right) \label{PCirSol},
\end{eqnarray} 
where the global sign $(+)$ is for the counterclockwise motion and the $(-)$ sign for the motion in the clockwise 
direction. The symbol sgn$(x)$ stands for the piecewise sign function which we define in the form
\begin{equation}
\mbox{sgn}\left[\mbox{cn} \left( k_E (x-x_0) , \frac{1}{k_E} \right) \right] = \left \{ 
\begin{array}{cc}
+1 & \mbox{if} \, \,  \, \, \, (4n-1)K \leq k_E (x-x_0) < (4n+1)K, \\
-1 & \mbox{if}  \, \,  \, \, \, (4n+1)K \leq k_E (x-x_0) < (4n+3)K,
\end{array}
\right.
\end{equation}
and its role is to shorten the period of the function sn$( k_E (x-x_0) , 1/k_E )$ by half. This fact 
is in agreement with the expression for  $p(x)$ because the period of the elliptic function  
dn$( k_E (x-x_0) , 1/k_E )$ is $2K/k_E$ instead of $4K/k_E$, which is the period of the elliptic function  
sn$( k_E (x-x_0) , 1/k_E )$. The square modulus $k^2$ of the elliptic functions is equal to the inverse of the energy 
parameter $0<k^2=1/k_E^2<1$.

These are all the possible motions of the simple pendulum. It is straightforward to check that the solutions satisfy the equation of  conservation of energy (\ref{DimensionlessE}) by using the following relations between the Jacobi functions (in these relations the modulus satisfies $0<k^2 <1$) and its analogous relation for hyperbolic functions (which are obtained in the limit case $k=1$)

\begin{eqnarray}
\mbox{sn}^2(x,k) + \mbox{cn}^2(x,k) &=& 1, \label{relJac1}\\
\tanh^2(x)+\mbox{sech}^2(x) &=&1, \\
 k^2 \mbox{sn}^2(x,k)  + \mbox{dn}^2(x,k)  &=& 1. \label{relJac2}
\end{eqnarray}

\subsection{Imaginary time pendulum}\label{ImagTimePen}

In the analysis above, time was considered a real variable, and therefore in the solutions of the simple pendulum  only the real quarter period $K$ appeared. But Jacobi elliptic functions are defined in $\mathbb{C}$ and, for instance, the function $\mbox{sn}(z,k)$ of square modulus $0 < k^2 < 1$, besides the real primitive period $4K$, owns a pure imaginary primitive period $2iK_c$,  where $K$ and $K_c$ are  the so called quarter periods (see for instance \cite{Lawden}). In 1878 Paul Appell clarified the physical meaning of the imaginary time and the imaginary period in the 
oscillatory solutions of the pendulum \cite{Appell,Armitage}, by introducing an ingenious trick, he reversed the direction 
of the gravitational field: $g \rightarrow -g$,  i.e. now the gravitational field is upwards. In order the Newton equations of 
motion remain invariant under this change in the force, we must replace the real time variable $t$ by a purely 
imaginary one: $\tau \equiv  \pm it$. Implementing these changes in the equation of motion (\ref{eqmotion}) leads to 
the equation
\begin{equation}\label{NewImag}
\frac{d^2 \theta}{d\tau^2} - \frac{g}{r} \sin \theta =0.
\end{equation}
Writing this equation in dimensionless form requires the introduction of the pure imaginary time variable
$y \equiv  \pm \tau \sqrt{g/r}=\pm ix$. Integrating once the resulting dimensionless equation of motion gives origin to 
the following equation
\begin{equation}\label{ConsEner2}
\frac{1}{4} \left( \frac{d \theta}{dy} \right)^2 - \sin^2 \left( \frac{\theta}{2}\right) = -k_E^2,
\end{equation}
which looks like equation (\ref{DimensionlessE}) but with an inverted potential. In order to solve  this equation, we 
should flip the sign in the potential and rewrite the equation in such a way that it looks similar to equation 
(\ref{DimensionlessE}). To achieve this aim we start by shifting the value of the potential energy one unit such that its 
minimum value be zero. Adding a unit of energy to both sides of the equation leads to
\begin{equation}\label{ConsEneryIm}
\left(\frac{\mathbb{P}}{2} \right)^2 + \cos^2 \left( \frac{\theta}{2}\right) = 1-k_E^2.
\end{equation}
Here $\mathbb{P}$ is the momentum as function of imaginary time. The second step is to rewrite the potential energy 
in such a form it coincides with the potential energy of (\ref{DimensionlessE}) and, in this way, allowing us to compare solutions. We can accomplish this by a simple translation of the graph, for instance, by translating it an angle of $\pi/2$ 
to the right. Defining $\theta'= \theta - \pi$, we obtain
\begin{equation}\label{ConsEner4}
\left(\frac{\mathbb{P}}{2} \right)^2 + \sin^2 \left( \frac{\theta'}{2}\right) = 1-k_E^2.
\end{equation}
It is clear that $0< 1-k_E^2 <1$ for oscillatory motions and $1-k_E^2 <0$ for the circulating ones. It is not the purpose 
of this paper to review the whole set of solutions with imaginary time. The reader interested in the detailed construction 
can see \cite{Linares} for instance.

\section{The pendulum from the extended rigid body ($c\neq 0$ and $d \neq 0$)}\label{Buena}

In this section we review the relation between the rigid body and the simple pendulum as originally discussed by Holm and Marsden \cite{Marsden1}, and we extend it to pendulums of imaginary time. In every case we discuss the geometrical and the algebraic characteristics of the relations. 

A lesson learned from \cite{Marsden1} is that, in order to have the pendulum from the rigid body it is necessary that the 
$SL(2,\mathbb{R})$ transformation of the Casimir functions lead to new ones that geometrically represent 
two perpendicular elliptic cylinders.  In our parameterization, the mathematical conditions for this to happen is that
\begin{equation}\label{CylinderCond}
ae_i+b=0, \hspace{0.5cm} \mbox{and}  \hspace{0.5cm} ce_j+d=0, 
\hspace{0.5cm} \mbox{with}  \hspace{0.5cm} i \neq j.
\end{equation} 
However, due to the fact that the inertia parameters are not necessarily positive, these conditions are not attached to 
elliptical cylinders only, but also to hyperbolic cylinders  \cite{delaCruz:2017qkh}. In order to have control over all 
different possibilities to get pendulums from the rigid body it is necessary to list all different $SL(2,\mathbb{R})$ 
transformations that fulfill conditions  (\ref{CylinderCond}). The transformations are divided into three general sets, 
where each set contains different fixed  forms of the $SL(2,\mathbb{R})$ matrices. The three sets are determined by 
the conditions: i) $c\neq 0$ and $d \neq 0$. ii)  $c\neq 0$ and $d = 0$ and iii) $c=0$ and $d \neq 0$ \cite{David1992}. 
In the latter case a generic $SL((2,\mathbb{R})$ group element has the form
\begin{equation}
g= \left( 
\begin{array}{cc}
a & b \\
0 & 1/a
\end{array}
\right), \hspace{0.5cm} \mbox{with}  \hspace{0.2cm} b \in \mathbb{R}.
\end{equation}
We conclude that the surface ${\cal N}$ is a unitary sphere and therefore in this case it is not possible to degenerate 
the surface to a cylinder. As a consequence the pendulum can not arise from group elements of this kind and we 
must analyze only the cases i) and ii). Because the case discussed by Holm and Marsden belongs to the first set of 
conditions, we analyze it in this section and leave the set: $c\neq 0$ and $d = 0$ for section \ref{CNeq0D0}.

The first step in our analysis is to fully classify the different cases belonging to the set $c \neq 0$ and $d \neq 0$ 
in which the Casimir surfaces ${\cal H}$ and ${\cal N}$ as defined in equation (\ref{TransSL2}) fulfill conditions 
(\ref{CylinderCond}). It is clear that there are 6 different forms to satisfy these conditions. However 
these are not all independent, in fact, the difference between conditions: $ae_i+b=0$ and $ce_j+d=0$, with respect to 
conditions: $ae_j+b=0$ and $ce_i+d=0$, is that they interchange the role of the Casmir functions $H$ and $N$. Thus 
we can restrict ourselves to the analysis of the three cases (\ref{CylinderCond}) for $i<j$. Because solutions 
(\ref{Solutions1}) and (\ref{Solutions2}) depend on the relative value of the parameter $e_0$ respect to $e_2$, the 
classification of geometries for both ${\cal H}$ and ${\cal N}$ in general also depends on this relative value  
\cite{delaCruz:2017qkh}. This fact increases the number of cases to five. We show in table \ref{table1Comb} the 
different possible geometries for the Casimir $H$ and in table \ref{Bianchicd} the corresponding ones for the Casimir 
$N$. It is important to stress that in this classification we are using the fact that the space 
$\{u_1,u_2,u_3\}$ is $\mathbb{R}^3$.
\begin{table}[htb]  
\begin{tabular}{|c | c | c | c | c |} 
\hline
Situation & $ae_1+b$ & $ae_2+b$ & $ae_3+b$ & ${\cal H}$ surface \\
\hline 
\hline
1 &  $=0$ &  $< 0$ & $<0$ &  elliptic cylinder around $u_1$\\
2 & $>0$ &  $=0$ &  $<0$ & hyperbolic cylinder around $u_2$ with focus on $u_1$ $(e_2< e_0)$ \\
3 & $>0$ &  $=0$ &  $<0$ & hyperbolic cylinder around $u_2$ with focus on $u_3$  $(e_0< e_2)$ \\
4 & $>0$ &  $> 0$ &  $=0$ &  elliptic cylinder around $u_3$ \\
 \hline
\end{tabular}
\caption{Classification of the four different geometries for the Casimir function $H$, in the asymmetric extended rigid  
body.}\label{table1Comb}
\end{table}

\begin{table}[htb]  
\centering
\begin{tabular}{| c | c | c | c | c | c | c | c | c |} 
\hline
Situation & $ce_1+d$ & $ce_2+d$ & $ce_3+d$ & ${\cal N}$ surface \\
\hline 
\hline
5 & $=0$ & $< 0$& $< 0$ & elliptic cylinder around $u_1$\\
6 & $>0$ & $=0$ & $<0$ & hyperbolic cylinder around $u_2$  with focus on $u_1$ $(e_2< e_0)$ \\
7 & $>0$ & $=0$ & $<0$ & hyperbolic cylinder around $u_2$ with focus on $u_3$  $(e_0< e_2)$\\
8 & $>0$ & $>0$ & $=0$ &  elliptic cylinder around $u_3$ \\
\hline
\end{tabular}
\caption{Classification of the four different geometries for the Casimir function $N$, in the asymmetric extended rigid 
body.}\label{Bianchicd}
\end{table}

It is clear the five different possibilities we are referring to correspond to the intersections: 1-6, 1-7, 1-8, 2-8 y 3-8. The one discussed by Holm and Marsden corresponds to the case of the intersection of two elliptical cylinders 1-8, however we have four more possibilities which  correspond to the intersection of an elliptical cylinder and a hyperbolic cylinder.  For the best of our knowledge these four cases have not been discussed in the literature; one of the aims of this 
section is to give them an interpretation as pendulums. At this point it is convenient to stress the two main differences 
between our analysis and the original discussion in \cite{Marsden1}. 1) The discussion of Holm and Marsden was 
given using  the moments of inertia (\ref{OrderI}) as parameters whereas here we are using a different 
parameterization, the one in terms of the dimensionless inertia parameters $\{e_i\}$, which produce slight differences 
in the discussion. 2) Additionally to the original momentum map $\{ u_i\} \rightarrow \{ u_i (\theta,p)\}$ whose main 
characteristic  is to have a real pendulum momentum $p$ or equivalently a real angular coordinate $\theta$ and a real 
time $t$, in our discussion we will consider also momentum maps where the real pendulum momentum $p$ is rewritten 
as $p=i \mathbb{P}$ with the new momentum: $\mathbb{P}=-ip$ purely imaginary. As discussed in subsection 
\ref{ImagTimePen} this momentum is a consequence of introducing a purely imaginary time. When time is real the final 
dimensionless angular momentum space  remains $\mathbb{R}^3$  whereas the associated pendulum phase space is 
$\mathbb{R} \times S^1$. When time is purely imaginary the final angular  momentum space is 
$\mathbb{R}^2 \times i \mathbb{R}$ whereas the pendulum phase space is built with a real coordinate $\theta$ and
a purely imaginary momentum $\mathbb{P}$. We start discussing the original case of Holm and Marsden (1-8).

\subsection{Intersection of two elliptic cylinders}\label{TwoElliptic}

Choosing the $a$, $b$, $c$ and $d$ nonvanishing values of the group element $g \in SL(2,\mathbb{R})$ to satisfy 
\begin{equation}
a e_1+b=ce_3+d=0,
\end{equation}
fix two of the three free parameters of $g$.  Substituting $b$ and $d$ in the expressions (\ref{TransSL2}) for the 
Casimir surfaces we obtain
\begin{eqnarray}
{\cal H}: && -\frac{a}{2} (e_1-e_2)  \, u_2^2 -\frac{a}{2} (e_1-e_3) \, u_3^2 = -\frac{a}{2}( e_1-e_0), \\
{\cal N}: && \, \, \, \, \, \frac{c}{2}( e_1-e_3) \, u_1^2 + \frac{c}{2} (e_2-e_3) \, u_2^2 = \,  \,\, \, \frac{c}{2} (e_0-e_3).
\end{eqnarray}
Notice that the surfaces do not depend of the specific values of both $a$ and $c$. However we can fix one of these parameters in terms of the other using the condition $ad-bc=1$, which in this case produce the condition $ac=1/(e_1-e_3)$. In other words, Holm and Marsden considered an element $g \in SL(2,\mathbb{R})$ of the type
\begin{equation}\label{gMarsden}
g= \left( 
\begin{array}{cc}
\frac{1}{c(e_1-e_3)} & -\frac{e_1}{c(e_1-e_3)} \\
c & -ce_3
\end{array}
\right),
\end{equation}
which depends only on one free parameter $c \neq 0$. Originally this parameter was fixed to the value $c=1$ although it 
is not necessary to fix it because the $SL(2,\mathbb{R})$ combination of $h$ and $l$ leads to expressions that do not 
depend on $c$. It is clear that given the ordering (\ref{esOrdering}), the coefficients in (\ref{gMarsden}) have the 
property $a>0$, $b<0$ and $d>0$.

\begin{figure}[h!]
\begin{center}
\subfigure[Rigid body solutions  \label{RgidBodyFig}] {\includegraphics[scale=0.4]{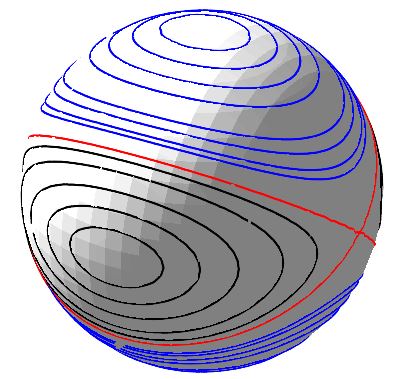}} \hfill 
\subfigure[Pendulum with Hamiltonian ${\cal H}$  \label{PendulumVer} ]{\includegraphics[scale=0.5]{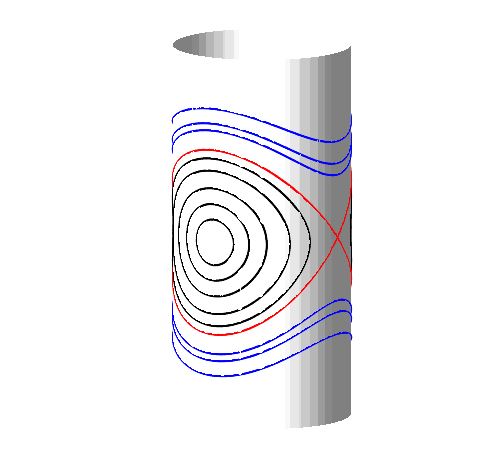}} \hfill 
\subfigure[Pendulum with Hamiltonian ${\cal N}$  \label{PendulumHor}]{\includegraphics[scale=0.5]{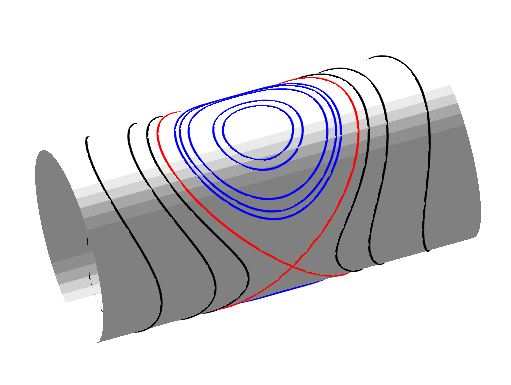} }  
\end{center}
\caption{Figure \ref{RgidBodyFig} shows the solution trajectories of the vector of angular momentum in the body-fixed frame, for the $SO(3)$ rigid body. The ones corresponding to $e_2<e_0$ are drawn in black whereas the ones corresponding to $e_0<e_2$ are drawn in blue. The red lines correspond to the separatrixes.  Fig. \ref{PendulumVer} corresponds to the phase space of a simple pendulum with Hamiltonian ${\cal H}$ and cotangent space ${\cal N}$. The oscillatory movements are drawn in black whereas the circulating ones are drawn in blue. The colors correspond to the ones in fig. \ref{RgidBodyFig} and represent the relation between them. Fig. \ref{PendulumHor} corresponds to the phase space of a simple pendulum with Hamiltonian ${\cal N}$ and cotangent space ${\cal H}$. The oscillatory trajectories are drawn in blue whereas the circulating ones are drawn in black. The colors correspond to the ones in fig. \ref{RgidBodyFig} and represent the relation between them.}
\label{GeometryFig}
\end{figure}

The equations that determine the Casimir surfaces can be written finally as
\begin{eqnarray}
{\cal H}: && k_2^2 \, u_2^2 + \, u_3^2 = \frac{e_1-e_0}{e_1-e_3}, \label{Hcal}\\
{\cal N}: &&  u_1^2 + k_1^2 \, u_2^2 = \frac{e_0-e_3}{ e_1-e_3} , \label{Ncal}
\end{eqnarray}
which generically represent two elliptic cylinders, ${\cal H}$ with axis on cordinate $u_1$ and ${\cal N}$ with 
axis along $u_3$. Here $k_1^2$ and $k_2^2$ are the ones defined in (\ref{defke}). At this point the transverse sections 
of the cylinders are ellipses whose semi-major and semi-minor axes depend on the value of the parameter $e_0$, 
i.e. they have variable size and the intersections are exactly the same as the ones in fig. \ref{RgidBodyFig} because 
the $SL(2,\mathbb{R})$ transformations change the geometry of the Casimirs but leave invariant the intersections 
\cite{Nambu:1973qe}. 

The strategy is now to implement a variable change in such a way that one of the cylinders becomes circular using 
only trigonometric functions. Once this aim is achieved, the second cylinder can not be mapped to a circular cylinder 
simultaneously without introducing elliptic functions. This strategy can be applied in two  different ways, in one case 
${\cal H}$ becomes the circular cylinder, whereas in the second case the Casimir whose geometry becomes the 
circular cylinder is  ${\cal N}$. Interestingly these two cases produce the same physical system as expected for a 
bi-hamiltonian system, although the physical origin differs in both cases as we will explain. The variable changes or 
technically the momentum maps we implement \cite{MarsdenRatiu}, are analogous to the one performed by Holm and 
Marsden. As a result of the mapping one of the resulting Casimir functions obtain the form of the Hamiltonian of the 
simple pendulum and we can identify directly if the energy produces an oscillating or a circulating movement. 
Additionally we also introduce momentum maps for imaginary time.\\

\noindent $\bullet$ Simple pendulum with real time and Hamiltonian ${\cal H}$: Consider the momentum map
\begin{equation}\label{OscMap}
u_1\equiv \sqrt{\frac{e_0-e_3}{e_1-e_3}} \cos \left( \frac{\theta}{2}\right) , \hspace{0.3cm}
u_2\equiv \frac{1}{k_1}\sqrt{\frac{e_0-e_3}{e_1-e_3}} \sin  \left( \frac{\theta}{2}\right),  \hspace{0.3cm}
u_3\equiv  \frac{k_2}{k_1} \sqrt {\frac{e_0-e_3}{e_1-e_3}}  \, \frac{p}{2}.
\end{equation}
The expression for the Casimir surface ${\cal N}$ is explicitly satisfied whereas the expression for the Casimir 
${\cal H}$ becomes
\begin{equation}\label{HOscReal}
{\cal H}: \sin^2  \left( \frac{\theta}{2}\right) + \left( \frac{p}{2} \right)^2= 
\frac{e_1-e_0}{e_0-e_3} \cdot  \frac{e_2-e_3}{e_1-e_2} \equiv m^2.
\end{equation}
${\cal N}$ is interpreted as the cotangent space and has the geometry of a circular cylinder of unitary radius. On the other side, according to equation (\ref{DimensionlessE}) ${\cal H}$ is the Hamiltonian of a pendulum. This is the 
analogous of the relation found by Holm and Marsden \cite{Marsden1}. Because situation 1 in table \ref{table1Comb} 
is valid for any value of $e_0$ in the interval $e_3 < e_0 < e_1$ we have that $0<m^2<1$ for $e_0 > e_2$, $m^2=1$ 
for $e_0=e_2$ and $m^2>1$ for $e_0<e_2$ (see table \ref{valuesSM}). Therefore for $m^2<1$ we have an oscillatory 
movement of the pendulum, for $m^2=1$ we have a critical movement and for $m^2>1$ we have a circulating one. 
Fig. \ref{PendulumVer} shows the complete phase space of the pendulum.

Regarding the generators (\ref{BasicGen}) of the system, it is straightforward to check they can be redefined as
\begin{eqnarray}
Y_{u_1}&\equiv& \frac{1}{c \alpha} X_{u_1}= -\frac{(e_1-e_3)}{\alpha} \, k^2_1\, u_2 \, \partial_3, \nonumber \\
Y_{u_2}&\equiv& \frac{k_1}{c \alpha} X_{u_2}= \frac{(e_1-e_3)}{\alpha} \, k_1 \, u_1 \partial_3, 
\label{HVFHM}\\
Y_{u_3}&\equiv& \frac{1}{c k_1 (e_1-e_3)} X_{u_3}= k_1 \, u_2 \, \partial_1 - \frac{1}{k_1} \, u_1 \, 
\partial_2, \nonumber
\end{eqnarray} 
which satisfy the Lie-Poisson algebra $ISO(2)$
\begin{equation}\label{LPAHMOsc}
[Y_{u_1},Y_{u_2}]=0, \hspace{0.5cm} [Y_{u_2},Y_{u_3}]=Y_{u_1},  \hspace{0.5cm}   [Y_{u_3},Y_{u_1}]=Y_{u_2}.
\end{equation}
A Lie-Poisson structure for the pendulum can be introduced on the cylindrical surface ${\cal N}=1$ 
\cite{Marsden1}, in terms of the variables $\theta, p$ by defining the Lie-Poisson bracket as 
\begin{equation}
\{ f,g \}_{\cal N} = - \nabla {\cal N} \cdot (\nabla f \times \nabla g).
\end{equation} 
An straightforward calculation gives 
\begin{equation}\label{LPSHHR}
\left \{ f,g \right \}_{\cal N}=-8\frac{(e_{1}-e_{3})(e_{2}-e_{3})}{\sqrt{(e_{1}-e_{2})(e_{0}-e_{3})}} 
\left(\frac{\partial f}{\partial \theta}\frac{\partial g}{\partial p}-\frac{\partial f}{\partial p}\frac{\partial g}{\partial \theta}\right),
\end{equation}
which shows that the variables $\theta$ and $p$ are canonically conjugate up to a scale factor:  
$\left \{ p,\theta \right \}_{\cal N}=8\frac{(e_{1}-e_{3})(e_{2}-e_{3})}{\sqrt{(e_{1}-e_{2})(e_{0}-e_{3})}}$. In terms of this 
bracket the canonical equations of motion are given by
\begin{equation}
\frac{d \theta}{d \tau} = \left \{ {\cal H},\theta \right \}_{\cal N} \hspace{0.5cm} \mbox {and}  \hspace{0.5cm} 
\frac{d p}{d \tau} = \left \{ {\cal H},p \right \}_{\cal N}.
\end{equation}
Combining these we obtain the Newton equation of motion for the simple pendulum
\begin{equation}\label{EOMHHR}
\frac{d^{2}\theta}{d\tau^{2}}=-16\frac{(e_{1}-e_{3})^{2}(e_{2}-e_{3})^{2}}{(e_{1}-e_{2})(e_{0}-e_{3})} \sin \theta.
\end{equation}
\\

\noindent $\bullet$ Simple pendulum with imaginary time and Hamiltonian ${\cal H}$: Consider the momentum map
\begin{equation}\label{OscMapIm}
u_1\equiv \sqrt{\frac{e_0-e_3}{e_1-e_3}} \sin \left( \frac{\theta'}{2}\right) , \hspace{0.3cm}
u_2\equiv \frac{1}{k_1}\sqrt{\frac{e_0-e_3}{e_1-e_3}} \cos  \left( \frac{\theta'}{2}\right),  \hspace{0.3cm}
u_3\equiv  i \frac{k_2}{k_1} \sqrt {\frac{e_0-e_3}{e_1-e_3}}  \, \frac{\mathbb{P}}{2}.
\end{equation}
The expression for the Casimir surface ${\cal N}$ is again explicitly satisfied whereas the form of the Casimir 
${\cal H}$ becomes
\begin{equation}\label{HOscImTime}
{\cal H}: \sin^2  \left( \frac{\theta'}{2}\right) + \left( \frac{\mathbb{P}}{2} \right)^2= m_c^2 ,
\end{equation}
where $m_c^2$ is the complementary modulus as defined in (\ref{equ}). The Casimir ${\cal H}$ 
represents the Hamiltonian of the simple pendulum with imaginary time (\ref{ConsEner4}). Again because 
(\ref{HOscImTime}) is valid for any value of $e_0$, we have all different movements of the pendulum. In particular 
for the oscillatory motions: $e_2<e_0$ and $0<m_c^2<1$, for the asymptotical motion: $e_0=e_2$ and $m_c^2=0$, 
whereas for the circulating movements: $e_0<e_2$ and $m^2_c<0$ (see table \ref{valuesSM}). Notice that the explicit 
presence of the imaginary number $i$ in coordinate $u_3$ geometrically plays the role of changing the elliptic cylinder 
into a hyperbolic cylinder although the former is defined in a real space $\mathbb{R}^3$ whereas the later is defined in 
a complex space $\mathbb{R}^2\times i \mathbb{R}$. Regarding the Lie-Poisson algebra it is clear that here we have 
the same $ISO(2)$ algebra (\ref{LPAHMOsc}). The two dimensional Lie-Poisson bracket in this case has a purely  
imaginary scale factor
\begin{equation}\label{2dPBCT}
\left \{ f,g \right \}_{\cal N}=-8i \frac{(e_{1}-e_{3})(e_{2}-e_{3})}{\sqrt{(e_{1}-e_{2})(e_{0}-e_{3})}} 
\left(\frac{\partial f}{\partial \theta}\frac{\partial g}{\partial \mathbb{P}}-
\frac{\partial f}{\partial  \mathbb{P}}\frac{\partial g}{\partial \theta}\right),
\end{equation}
which is in agreement with the fact that the Newton equation for the simple pendulum with imaginary time has an extra 
minus sign in the force term
\begin{equation}\label{NECT}
\frac{d^{2}\theta}{d\tau^{2}}=16\frac{(e_{1}-e_{3})^{2}(e_{2}-e_{3})^{2}}{(e_{1}-e_{2})(e_{0}-e_{3})} \sin \theta.
\end{equation}
Because the rigid body is a bi-Hamiltonian system it is interesting to obtain the simple pendulum but now having the 
Casimir function ${\cal N}$ as the hamiltonian and the Casimir function ${\cal H}$ as the cotangent space.\\

\noindent $\bullet$ Simple pendulum with real time and Hamiltonian ${\cal N}$: Consider the momentum map
\begin{equation}\label{CirMap}
u_1 \equiv  \frac{k_1}{k_2} \sqrt{ \frac{e_1-e_0}{e_1-e_3}} \, \frac{p}{2},  \hspace{0.3cm} 
u_2 \equiv \frac{1}{k_2} \sqrt{\frac{e_1-e_0}{e_1-e_3}} \sin \left( \frac{\theta}{2}\right), \hspace{0.3cm}
u_3 \equiv \sqrt{\frac{e_1-e_0}{e_1-e_3}} \cos \left( \frac{\theta}{2}\right).
\end{equation}
Now the expression for the Casimir function ${\cal H}$ is explicitly satisfied whereas the expression for the Casimir 
${\cal N}$ becomes
\begin{equation}\label{HamN} 
{\cal N}:  \left( \frac{p}{2} \right) ^2 + \sin^2\left( \frac{\theta}{2}\right) = \frac{1}{m^2}, 
\end{equation}
which according to equation (\ref{DimensionlessE}) corresponds to the energy of a simple pendulum. Because 
situation 8 in table \ref{Bianchicd} is valid for any value of $e_0$ in the interval $e_3<e_0<e_1$, in equation 
(\ref{HamN}) we have all different movements of the pendulum. It is worth to stress that whereas for values $e_0<e_2$ 
for which $0<m^2<1$ the movements in (\ref{HOscReal}) are oscillatory (see fig. \ref{PendulumVer}), in (\ref{HamN}) 
for the same values of $e_0$: $1<1/m^2< \infty$ and therefore we have circulating pendulums (see 
fig. \ref{PendulumHor}). Something analogous occur for the circulating movements in (\ref{HOscReal}) and the 
oscillatory ones in (\ref{HamN}). The asymptotical cases are given for $e_0=e_2$ for which $m^2=1$ in both 
(\ref{HOscReal}) and (\ref{HamN}).

Let us emphasize that the circular cylinder ${\cal H}$ has unitary radius. In this case the Hamiltonian vector field 
associated with the coordinates $u_i$'s are given by equations (\ref{BasicGenN}). Redefining these vector fields as
\begin{eqnarray}
\tilde{Y}_{u_1}&\equiv& \frac{1}{a (e_1-e_3) k_2} \tilde{X}_{u_1}= \frac{1}{k_2} \, u_3 \, \partial_2 - k_2 \, 
u_2 \partial_3, \nonumber \\
\tilde{Y}_{u_2}&\equiv& \frac{1}{a \alpha} \tilde{X}_{u_2} = -\frac{(e_1-e_3)}{\alpha} \, u_3 \, \partial_1, 
\label{HVFHMT}\\
\tilde{Y}_{u_3}&\equiv& \frac{1}{a \alpha k_2} \tilde{X}_{u_3}= \frac{e_1-e_3}{\alpha} \, k_2 \, u_2 \, \partial_1,
\nonumber
\end{eqnarray} 
they satisfy the $ISO(2)$ Lie-Poisson algebra
\begin{equation}\label{LPAHMTCir}
[\tilde{Y}_{u_1},\tilde{Y}_{u_2}]=\tilde{Y}_{u_3}, \hspace{0.5cm} [\tilde{Y}_{u_2},\tilde{Y}_{u_3}]= 0,  \hspace{0.5cm}   
[\tilde{Y}_{u_3},\tilde{Y}_{u_1}]=\tilde{Y}_{u_2}. 
\end{equation}
Regarding the Lie-Poisson bracket in terms of the variables ($\theta$, $p$) on the cylindrical surface ${\cal H}=1$,  
and defined as
\begin{equation}
\{ f,g \}_{\cal H} = - \nabla {\cal H} \cdot (\nabla f \times \nabla g), 
\end{equation} 
we have 
\begin{equation}\label{LPtpReal}
\left \{ f,g \right \}_{\cal H}=8\frac{(e_{1}-e_{3})(e_{1}-e_{2})}{\sqrt{(e_{1}-e_{0})(e_{2}-e_{3})}} 
\left(\frac{\partial f}{\partial \theta}\frac{\partial g}{\partial p}-\frac{\partial f}{\partial p}\frac{\partial g}{\partial \theta}\right).
\end{equation}
The Newton equation associated to this Lie-Poisson bracket is
\begin{equation}\label{NewRTHamN}
\frac{d^{2}\theta}{d\tau^{2}}=-16\frac{(e_{1}-e_{3})^{2}(e_{1}-e_{2})^{2}}{(e_{1}-e_{0})(e_{2}-e_{3})} \sin \theta.
\end{equation}
\\

\noindent $\bullet$ Simple pendulum with imaginary time and Hamiltonian ${\cal N}$: Consider the momentum map
\begin{equation}\label{CirMapIm}
u_1 \equiv i  \frac{k_1}{k_2} \sqrt{ \frac{e_1-e_0}{e_1-e_3}} \, \frac{\mathbb{P}}{2},  \hspace{0.3cm} 
u_2 \equiv \frac{1}{k_2} \sqrt{\frac{e_1-e_0}{e_1-e_3}} \cos \left( \frac{\theta'}{2}\right), \hspace{0.3cm}
u_3 \equiv \sqrt{\frac{e_1-e_0}{e_1-e_3}} \sin \left( \frac{\theta'}{2}\right).
\end{equation}
The expression for the Casimir function ${\cal H}$ is explicitly satisfied whereas the expression for the Casimir 
${\cal N}$ becomes
\begin{equation}\label{NCirImTime}
{\cal N}:  \left( \frac{\mathbb{P}}{2} \right) ^2 + \sin^2\left( \frac{\theta'}{2}\right) =- \frac{m_c^2}{m^2}.
\end{equation}
This is the Hamiltonian of a simple pendulum with imaginary time. As in the previous cases, because (\ref{NCirImTime}) is valid for any value of $e_0$, it includes all different movements of the pendulum. Because the cotangent space is ${\cal H}$, the $ISO(2)$ Lie-Poisson algebra is given by (\ref{LPAHMTCir}), and the Lie-Poisson bracket in terms of the 
variables ($\theta$, $ \mathbb{P}$) is similar to (\ref{LPtpReal}) and differs only by a factor of $i$ in the scale factor.\\

The geometrical interpretation of these results is straightforward.  If we start with a specific $SO(3)$ rigid 
body, the three moments of inertia $\{I_i\}$ are given, which fix the parameter $\kappa$ and therefore the three 
dimensionless parameters of inertia $\{e_i\}$.  The different solutions are obtained for different values of $e_0$. After 
the $SL(2,\mathbb{R})$ gauge transformation of the Casimir surfaces, for the momentum maps 
(\ref{OscMap}) and (\ref{OscMapIm}) we obtain a unitary circular cylinder ${\cal N}$ with the axis along the $u_3$ 
direction representing the cotangent space. On the other side, geometrically ${\cal H}$ is 
an element of a set of elliptic-(hyperbolic) cylinders with axis along the $u_1$ direction which physically are energy 
surfaces, because ${\cal H}$ is the Hamiltonian of the simple pendulum (see equations (\ref{DimensionlessE}) and  
(\ref{ConsEner4})). Intersections of the single circular cylinder ${\cal N}$ with the set of elliptic-(hyperbolic) cylinders 
${\cal H}$ represent all the movements of the pendulum. On the other side, for the momentum maps (\ref{CirMap}) and 
(\ref{CirMapIm}) the surface ${\cal H}$ becomes an unitary circular cylinder with axis along the $u_1$ direction 
representing the cotangent space and ${\cal N}$ is an element of a set of elliptic-(hyperbolic) cylinders that represent 
the Hamiltonian of the simple pendulum. 

\subsection{Intersection of a hyperbolic cylinder and an elliptic cylinder I}\label{OscPenIntHypEllip}

\begin{figure}[h!]
{\includegraphics[scale=0.4]{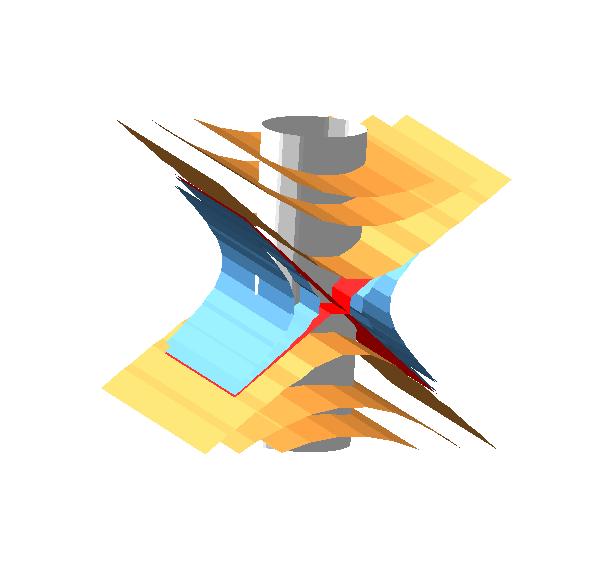}} \hfill 
\caption{Figure shows the geometry of the hyperbolic cylinders for different values for $e_0$. The geometry 
corresponding to $e_2<e_0$ is drawn in blue whereas the ones corresponding to $e_0<e_2$ are drawn in yellow. The 
red planes correspond to the separatrixes.}
\label{GeometryFig2}
\end{figure}

Let us consider the cases 2-8 and 3-8, which correspond to the intersection of a hyperbolic cylinder and an elliptical 
cylinder. In these cases the $SL(2,\mathbb{R})$ group element has the form
\begin{equation}\label{gImag}
g= \left( 
\begin{array}{cc}
\frac{1}{c(e_2-e_3)} & -\frac{e_2}{c(e_2-e_3)} \\
c & -ce_2
\end{array}
\right).
\end{equation}
As a consequence the Casimir surfaces  are given by the expressions
\begin{eqnarray}
{\cal H}: && k_2^2 \, u_1^2 -  k_1^2\, u_3^2 = \frac{e_0-e_2}{e_1-e_3} , \label{Hcal2-8}\\
{\cal N}: && u_1^2 + k_1^2 \, u_2^2 = \frac{e_0-e_3}{e_1-e_3}. \label{Ncal2-8}
\end{eqnarray}
Notice that the Casimir ${\cal N}$ is exactly the same as (\ref{Ncal}), as it should be since in both cases we are dealing 
with the situation 8 (see table \ref{Bianchicd}), which represents an elliptic cylinder of unitary radius with axis along 
$u_3$. The cases 2-8 and 3-8 differ in the sign of the right hand side of the Casimir ${\cal H}$ (\ref{Hcal2-8}), 
for the case 2-8 we have $e_0 - e_2>0$, whereas for the case 3-8 we have $e_0 - e_2<0$. Geometrically the different 
signs change the orientation of the hyperbolic cylinders (see fig \ref{GeometryFig2}).  Again we expect to have two 
different momentum maps, one in which ${\cal N}$ is the cotangent space and ${\cal H}$ is the Hamiltonian and a 
second one where  ${\cal H}$ is the cotangent space and ${\cal N}$ is the Hamiltonian. \\

\noindent $\bullet$ Simple pendulum with real time and Hamiltonian ${\cal H}$: As we have discussed previously, 
under the momentum map (\ref{OscMap})
\begin{equation*}
u_1\equiv \sqrt{\frac{e_0-e_3}{e_1-e_3}} \cos  \left( \frac{\theta}{2}\right), \hspace{0.3cm}
u_2\equiv  \frac{1}{k_1} \sqrt{\frac{e_0-e_3}{e_1-e_3}} \sin  \left( \frac{\theta}{2}\right),  \hspace{0.3cm}
u_3\equiv  \frac{k_2}{k_1} \sqrt{\frac{e_0-e_3}{e_1-e_3}} \, \left( \frac{p}{2} \right),
\end{equation*}
the expression for the Casimir ${\cal N}$ is automatically satisfied.  On the other side the Casimir surface 
(\ref{Hcal2-8}) coincides with the Hamiltonian (\ref{DimensionlessE}) and (\ref{HOscReal}) 
\begin{equation*}\label{HReal2}
{\cal H}: \sin^2  \left( \frac{\theta}{2}\right) + \left( \frac{p}{2} \right)^2 = m^2 ,
\end{equation*}
where $0<m^2<1$ for the case 2-8 ($e_0 - e_2>0$) and therefore represent the oscillatory movements, whereas 
$1<m^2< \infty$ for the case 3-8  ($e_0 - e_2<0$) which represent the circulating movements. As usual the case 
$e_0=e_2$ produce the separatrix. Regarding the Hamiltonian vector fields associated to the coordinates, they are 
the same as (\ref{HVFHM}) and therefore they satisfy the Lie-Poisson algebra $ISO(2)$ (\ref{LPAHMOsc}). A 
straightforward calculation gives the two dimensional Lie-Poisson bracket (\ref{LPSHHR}) and the Newton equation 
(\ref{EOMHHR}). These results show that this case and IV-A represent the same dynamics, although they come from 
a different geometry for the Casimir ${\cal H}$.\\

\noindent $\bullet$ Simple pendulum with imaginary time and Hamiltonian ${\cal H}$: As in section \ref{TwoElliptic}, 
under the momentum map
\begin{equation}
u_1\equiv \sqrt{\frac{e_0-e_3}{e_1-e_3}} \sin  \left( \frac{\theta'}{2}\right), \hspace{0.3cm}
u_2\equiv  \frac{1}{k_1} \sqrt{\frac{e_0-e_3}{e_1-e_3}} \cos  \left( \frac{\theta'}{2}\right),  \hspace{0.3cm}
u_3\equiv i \frac{k_2}{k_1} \sqrt{\frac{e_0-e_3}{e_1-e_3}} \, \left( \frac{\mathbb{P}}{2} \right),
\end{equation}
the expression for the Casimir ${\cal N}$ is automatically satisfied whereas the Casimir ${\cal H}$ becomes
\begin{equation}\label{HImag2}
{\cal H}: \sin^2  \left( \frac{\theta'}{2}\right) + \left( \frac{\mathbb{P}}{2} \right)^2 = m_c^2.
\end{equation}
where $-\infty < m_c^2<1$.  Notice that the geometrical role of writing the coordinate $u_3$ in terms of an imaginary momentum $\mathbb{P}$ is to map the hyperbolic cylinder ${\cal H}$ to an elliptic cylinder with axis along 
the $u_2$ direction. According to equation (\ref{ConsEner4}), ${\cal H}$ represents the Hamiltonian for a simple 
pendulum of imaginary time and energy $m^2=1-m_c^2$, as the one in (\ref{HReal2}). As expected, the two 
dimensional Lie-Poisson structure and Newton equations coincide with (\ref{2dPBCT}) and (\ref{NECT}).\\

Next natural mapping is the one that maps the rigid body system to a cotangent space ${\cal H}$, with hamiltonian 
${\cal N}$. By inspection of the Casimir (\ref{Hcal2-8}), this can be achieved introducing a purely imaginary coordinate. 
Doing this allows to change the hyperbolic nature of the cylinders defined in $\mathbb{R}^3$ to elliptic cylinders 
defined in $\mathbb{R}^2 \times i\mathbb{R}$.\\

\noindent $\bullet$ Circulating simple pendulum with real time, imaginary coordinate and Hamiltonian ${\cal N}$: 
Consider the momentum map
\begin{equation}\label{CSPRTICN}
u_1\equiv  \frac{1}{k_2}\sqrt{\frac{e_0-e_2}{e_1-e_3}} \sin  \left( \frac{\theta}{2}\right), \hspace{0.3cm}
u_2\equiv  \frac{1}{k_1 k_2}\sqrt{\frac{e_0-e_2}{e_1-e_3}}  \frac{p}{2},  \hspace{0.3cm}
u_3\equiv  \frac{i}{k_1}\sqrt{\frac{e_0-e_2}{e_1-e_3}} \cos  \left( \frac{\theta}{2}\right).
\end{equation}
Here we are assuming: $e_0-e_2>0$ (situation 2 in table \ref{table1Comb}). For such a map the Casimir ${\cal H}$ is 
automatically satisfied whereas the Casimir ${\cal N}$ takes the form
\begin{equation}
{\cal N}: \sin^2  \left( \frac{\theta}{2}\right) + \left( \frac{p}{2} \right)^2 = \frac{1}{m_c^2},
\end{equation}
where $1< 1/m_c^2 < \infty$ (see table \ref{valuesSM}) and therefore the Hamiltonian equation represents only 
circulating movements of the simple pendulum. 

Because the Hamiltonian is given by ${\cal N}$, the suitable Hamiltonian vector fields associated to the coordinates are 
given by equation (\ref{BasicGenN}). Redefining them as
\begin{eqnarray}
\tilde{Y}_{u_1}&\equiv& \frac{k_2}{a \beta k_1} \tilde{X}_{u_1}= \frac{1}{\beta} \sqrt{(e_1-e_2)(e_2-e_3)} \, u_3 
\partial_2,
\nonumber \\
\tilde{Y}_{u_2}&\equiv& \frac{1}{a \sqrt{(e_1-e_3)(e_2-e_3)}} \tilde{X}_{u_2}= -\frac{k_2}{k_1} u_1 \partial_2 - 
\frac{k_1}{k_2} u_3 \partial_1,  \label{HVFHMTIT} \\
\tilde{Y}_{u_3}&\equiv& \frac{1}{a \beta} \tilde{X}_{u_3} = \frac{1}{\beta} (e_1-e_2)  \, \, u_1 \partial_2, \nonumber
\end{eqnarray} 
they satisfy the $ISO(1,1)$ Lie-Poisson algebra 
\begin{equation}\label{LPAHMT}
[\tilde{Y}_{u_1},\tilde{Y}_{u_2}]=\tilde{Y}_{u_3}, \hspace{0.5cm} [\tilde{Y}_{u_2},\tilde{Y}_{u_3}]= -\tilde{Y}_{u_1},  
\hspace{0.5cm}   [\tilde{Y}_{u_3},\tilde{Y}_{u_1}] = 0.
\end{equation}
Regarding the two-dimensional Lie-Poisson bracket, it has the form
\begin{equation}
\left \{ f,g \right \}_{\cal N}=-  8 i\frac{(e_{1}-e_{2})(e_{2}-e_{3})}{\sqrt{(e_{0}-e_{2})(e_{1}-e_{3})}}
\left(\frac{\partial f}{\partial \theta}\frac{\partial g}{\partial p}-\frac{\partial f}{\partial p}\frac{\partial g}{\partial \theta}\right),
\end{equation}
while the equation of motion goes as:
\begin{equation}\label{EOMHNImghyp}
\frac{d^{2}\theta}{d\tau^{2}}=16\frac{(e_{2}-e_{3})^{2}(e_{1}-e_{2})^{2}}{(e_{1}-e_{3})(e_{0}-e_{2})} \sin \theta,
\end{equation}
At this point the sign in the Newton equation seems to be wrong, but this sign is reflecting the fact that we have 
applied a complex map in coordinate $u_3$. 

Although we have presented the analysis assuming $e_2-e_0<0$, it is worthing to point out that the same physical 
situation emerge also for the case $e_2-e_0>0$. The only difference is to introduce a mapping where the complex 
coordinate is $u_1$ instead of $u_3$. We get as a conclusion of this subsection that, for the situation where the 
intersection of Casimir functions is among a hyperbolic cylinder ${\cal H}$ and a elliptic cylinder ${\cal N}$, if the 
Hamiltonian is given by ${\cal H}$ and the cotangent space is given by ${\cal N}$ then we get the whole motions of the 
simple pendulum and the Lie algebra of the extended rigid body is $ISO(2)$, but in the case where ${\cal H}$ is the 
Hamiltonian and ${\cal H}$ is the cotangent space we only get the circulating motions and the Lie algebra is 
$ISO(1,1)$.\\

\noindent $\bullet$ Circulating simple pendulum with imaginary time, imaginary coordinate and Hamiltonian ${\cal N}$: 
Consider the mapping 
\begin{equation}
u_1\equiv  \frac{1}{k_2}\sqrt{\frac{e_0-e_2}{e_1-e_3}} \cos  \left( \frac{\theta'}{2}\right), \hspace{0.3cm}
u_2\equiv  \frac{i}{k_1 k_2}\sqrt{\frac{e_0-e_2}{e_1-e_3}}  \frac{\mathbb{P}}{2},  \hspace{0.3cm} \label{ProbMP}
u_3\equiv  \frac{i}{k_1}\sqrt{\frac{e_0-e_2}{e_1-e_3}} \sin  \left( \frac{\theta'}{2}\right).
\end{equation}
For such a map, where we are assuming $e_0-e_2>0$, the Casimir ${\cal H}$ is automatically satisfied whereas the 
Casimir ${\cal N}$ takes the form
\begin{equation}
{\cal N}: \left( \frac{\mathbb{P}}{2} \right)^2  +  \sin^2  \left( \frac{\theta'}{2}\right) = - \frac{m^2}{m_c^2}.
\end{equation}
It is straightforward to notice that $-m^2/m_c^2<0$, and therefore we have as expected only circulating movements. This case also comes from the extend rigid body with Lie algebra $SO(1,1)$. Regarding the two dimensional 
Lie-Poisson bracket it has the following expression
\begin{equation}
\left \{ f,g \right \}_{\cal N}=  8 \frac{(e_{1}-e_{2})(e_{2}-e_{3})}{\sqrt{(e_{0}-e_{2})(e_{1}-e_{3})}} 
\left(\frac{\partial f}{\partial \theta}
\frac{\partial g}{\partial \mathbb{P}}-\frac{\partial f}{\partial \mathbb{P}}\frac{\partial g}{\partial \theta}\right),
\end{equation}
while the Newton equation of motion goes as (\ref{EOMHNImghyp}) but with an extra minus sign in the right hand side of the equation. Again the discordance of the Newton equation with respect to equation (\ref{ConsEner4}) is understood by the fact that the momentum map (\ref{ProbMP}) considers a complex transformation in coordinate $u_3$.

\subsection{Intersection of a hyperbolic cylinder and an elliptic cylinder II}\label{SecDeg}

As final cases we study the intersections between the elliptic cylinder and hyperbolic cylinder 1-6 and 1-7, for which the 
$SL(2,\mathbb{R})$ group elements take the form
\begin{equation}\label{gFinal}
g= \left( 
\begin{array}{cc}
\frac{1}{c(e_1-e_3)} & -\frac{e_1}{c(e_1-e_3)} \\
c & -ce_2
\end{array}
\right).
\end{equation}
For these intersections the expressions for the Casimir functions are 
\begin{eqnarray}
{\cal H}: && k_2^2 \, u_2^2 + \, u_3^2 = \frac{e_1-e_0}{e_1-e_3} , \label{Hcal1-7}\\
{\cal N}: && - k_2^2 \, u_1^2 + k_1^2 \, u_3^2 = \frac{e_2-e_0}{e_1-e_3}. \label{Ncal1-7}
\end{eqnarray}
The surface ${\cal H}$ has the same expression as (\ref{Hcal}) as it should be since we are working with the situation 
1 of table \ref{table1Comb}, which corresponds to an elliptic cylinder.\\

\noindent $\bullet$ Simple pendulum with real time and Hamiltonian ${\cal N}$: Consider the momentum map 
(\ref{CirMap})
\begin{equation*}
u_1\equiv \frac{k_1}{k_2} \sqrt{\frac{ e_1-e_0}{e_1-e_3} } \, \frac{p}{2}, \hspace{0.3cm}
u_2\equiv \frac{1}{k_2} \sqrt{\frac{e_1-e_0}{e_1-e_3}} \sin  \left( \frac{\theta}{2}\right),  \hspace{0.3cm}
u_3\equiv \sqrt{\frac{e_1-e_0}{e_1-e_3}} \cos \left( \frac{\theta}{2}\right) .
\end{equation*}
Clearly the Casimir ${\cal H}$ is automatically satisfied whereas the Casimir 
\begin{equation}
{\cal N}: \left( \frac{p}{2} \right)^2 +  \sin^2  \left( \frac{\theta}{2}\right)  = \frac{1}{m^2} .
\end{equation}
For values of $e_0 <e_2$, the quotient $1/m^2<1$ and therefore ${\cal N}$ represents the Hamiltonian for the simple 
pendulum in oscillatory motion whereas for $e_0>e_2$,  $1/m^2>1$ and ${\cal N}$ represents the Hamiltonian for the 
simple pendulum in circulating motion (\ref{DimensionlessE}). For the limiting case $e_0=e_2$, $1/m^2=1$ and we 
obtain the asymptotical motion.

As expected, in this case the two dimensional Lie-Poisson bracket coincides with (\ref{LPtpReal}) and therefore the 
Newton equation of motion is given by (\ref{NewRTHamN}). \\

\noindent $\bullet$ Simple pendulum with imaginary time and Hamiltonian ${\cal N}$: 
Consider the momentum map (\ref{CirMapIm})
\begin{equation}
u_1\equiv i \frac{k_1}{k_2} \sqrt{\frac{ e_1-e_0}{e_1-e_3} } \, \frac{\mathbb{P}}{2}, \hspace{0.3cm}
u_2\equiv \frac{1}{k_2} \sqrt{\frac{e_1-e_0}{e_1-e_3}} \cos  \left( \frac{\theta'}{2}\right),  \hspace{0.3cm}
u_3\equiv \sqrt{\frac{e_1-e_0}{e_1-e_3}} \sin \left( \frac{\theta'}{2}\right) .
\end{equation}
Under this map the Casimir ${\cal H}$ is automatically satisfied whereas the Casimir ${\cal N}$ becomes
\begin{equation}
{\cal N}: \left( \frac{\mathbb{P}}{2} \right)^2 +  \sin^2  \left( \frac{\theta'}{2}\right)  = -  \frac{m_c^2}{m^2} ,
\end{equation}
which is precisely the Hamiltonian of simple pendulum with imaginary time (\ref{NCirImTime}). Because we are using 
the same cotangent space (\ref{Hcal}) and (\ref{Hcal1-7}), the rest of physical quantities such as the two dimensional 
Lie-Poisson bracket and the Newton equation of motion coincide with the ones mentioned below equation  
(\ref{NCirImTime}).

Following the previous cases, next natural situation is the one that maps the rigid body system to a cotangent space 
${\cal N}$, with Hamiltonian ${\cal H}$. By inspection of Casimir (\ref{Ncal1-7}), this can be achieved introducing a 
momentum map with a purely imaginary coordinate.\\

\noindent $\bullet$ Circulating pendulum with real time, imaginary coordinate and Hamiltonian ${\cal H}$: Consider the 
momentum map
\begin{equation}\label{MapImu1}
u_1\equiv  \frac{i}{k_2} \sqrt{\frac{ e_2-e_0}{e_1-e_3} }  \cos  \left( \frac{\theta}{2}\right) , \hspace{0.3cm}
u_2\equiv \frac{1}{k_1 k_2} \sqrt{\frac{e_2-e_0}{e_1-e_3}} \, \frac{p}{2},  \hspace{0.3cm}
u_3\equiv \frac{1}{k_1}\sqrt{\frac{e_2-e_0}{e_1-e_3}} \sin \left( \frac{\theta}{2}\right) .
\end{equation}
Here we are assuming $e_2-e_0>0$ (situation 7 in table \ref{Bianchicd}).

Under this map the Casimir ${\cal N}$ is satisfied, while the Hamiltonian $\cal H$ has the following expression
\begin{equation}
{\cal H}: \left( \frac{p}{2} \right)^2 +  \sin^2  \left( \frac{\theta}{2}\right)  = -  \frac{m^2}{m_c^2} .
\end{equation}
For values $e_2-e_0>0$ the quotient $1<- m^2/m_c^2<\infty$ (see table \ref{valuesSM}) and we have a simple 
pendulum in circulating motion. Regarding the two dimensional Lie-Poisson bracket it has the following expression
\begin{equation}
\left \{ f,g \right \}_{\cal N}=  8i \frac{(e_{1}-e_{2})(e_{2}-e_{3})}{\sqrt{(e_{2}-e_{0})(e_{1}-e_{3})}}
\left(\frac{\partial f}{\partial \theta}\frac{\partial g}{\partial p}-\frac{\partial f}{\partial p}\frac{\partial g}{\partial \theta}\right),
\end{equation}
whereas the Newton equation of motion reads
\begin{equation}\label{EOMHNImghyp4}
\frac{d^{2}\theta}{d\tau^{2}}=16\frac{(e_{1}-e_{2})^{2}(e_{2}-e_{3})^{2}}{(e_{2}-e_{0})(e_{1}-e_{3})} \sin \theta.
\end{equation}
As the analogous physical situation in section \ref{OscPenIntHypEllip}, the extra minus sign that appears in 
the Newton equation (\ref{EOMHNImghyp4}) is due to the complex nature of the map (\ref{MapImu1}). 
As for the case $e_0>e_2$, we can go through it but the physical result will be the same, we only get the 
circulating movements of the pendulum.\\

\noindent $\bullet$ Circulating pendulum with imaginary time, imaginary coordinate and Hamiltonian ${\cal H}$: 
Consider the momentum map in which we assume $e_2>e_0$
\begin{equation}\label{LastMap}
u_1\equiv  \frac{i}{k_2} \sqrt{\frac{ e_2-e_0}{e_1-e_3} }  \sin  \left( \frac{\theta'}{2}\right) , \hspace{0.3cm}
u_2\equiv \frac{i}{k_1 k_2} \sqrt{\frac{e_2-e_0}{e_1-e_3}} \, \frac{\mathbb{P}}{2},  \hspace{0.3cm}
u_3\equiv \frac{1}{k_1}\sqrt{\frac{e_2-e_0}{e_1-e_3}} \cos\left( \frac{\theta'}{2}\right) .
\end{equation}
Under this map the Casimir $\cal N$ is satisfied, while the Hamiltonian $\cal H$ has the following expression
\begin{equation}
{\cal H}: \left( \frac{\mathbb{P}}{2} \right)^2 +  \sin^2  \left( \frac{\theta'}{2}\right)  =   \frac{1}{m_c^2} .
\end{equation}
It is clear that $-\infty < 1/m_c^2 <0$ and therefore as expected we only have circulating motions.

For completeness, the two dimensional Lie-Poisson bracket is
\begin{equation}
\left \{ f,g \right \}_{\cal N}=  -8 \frac{(e_{1}-e_{2})(e_{2}-e_{3})}{\sqrt{(e_{2}-e_{0})(e_{1}-e_{3})}} 
\left(\frac{\partial f}{\partial \theta}\frac{\partial g}{\partial \mathbb{P}}-\frac{\partial f}{\partial \mathbb{P}}\frac{\partial g}
{\partial \theta}\right),
\end{equation}
whereas the equation of motion reads
\begin{equation}\label{EOMHNImghyp5}
\frac{d^{2}\theta}{d\tau^{2}}=- 16\frac{(e_{1}-e_{2})^{2}(e_{2}-e_{3})^{2}}{(e_{2}-e_{0})(e_{1}-e_{3})} \sin \theta.
\end{equation}
Notice again the extra minus sign in (\ref{EOMHNImghyp5}) due to complex nature of the map (\ref{LastMap}).

\section{The pendulum from the extended rigid body ($c \neq 0$ and $d  = 0$)}\label{CNeq0D0}

Finally let us analyze the second general set of $SL(2,\mathbb{R})$ transformations. Regarding the conditions 
(\ref{CylinderCond}) the second one becomes modified to $ce_i=0$. Since in this case 
$c \neq 0$ then necessarily $e_i=0$. Even more, because we are restricted to the interval $\kappa \in (0,\pi/3)$ 
the only possibility is to have $e_2 = 0$ and because the restrictions  (\ref{CondDimless}) we also have 
$e_{1}=-e_{3}=\sqrt{3}/2$. We conclude that the conditions (\ref{CylinderCond}) can be satisfied only in two situations. 
In both of them $e_2=0$ and either $ae_1+b=0$ or $ae_3+b=0$. Due to the relation between $e_1$ and $e_3$ these 
two situations correspond to a transformation (\ref{TransSL2}) with different sign of the coefficient $a$. So without 
losing generality we can restrict ourselves to the case $a>0$ and analyze one of the situations. The second one can be 
obtained from the same formulas by changing the sign of $a$. In summary the conditions (\ref{CondDimless}) 
become
\begin{equation}
ae_1+b=0, \hspace{0.5cm} e_2=0,
\end{equation}
and a generic  $SL(2,\mathbb{R})$ element has the form
\begin{equation}\label{gFinaGen}
g= \left( 
\begin{array}{cc}
\frac{1}{ce_1} & -\frac{1}{c}\\
c & \, 0
\end{array}
\right).
\end{equation}
Notice that this transformation can be obtained from (\ref{gFinal}) modulo a factor of $1/2$ in the first arrow of the 
matrix, which does not change the geometry of the Casimir surface ${\cal H}$ and therefore  (\ref{gFinal}) corresponds 
to the $SL(2,\mathbb{R})$ transformation that takes the original Casimir surfaces to an elliptic and a hyperbolic 
cylinders. Notice that for $e_1=-e_3$ and $e_2=0$, the quotients (\ref{defke}) become equal $k_1^2=k_2^2=1/2$. 

It is clear that we do not have to developed this case further, since we can obtain it from a limiting case of the ones 
previously studied. Here we have two situations again, either $e_0<0$ (situation 1-7 of subsection 
\ref{SecDeg}) or $e_0>0$ (situation 1-6 of subsection \ref{SecDeg}). 

\section{Conclusions}\label{conclusions}

In this paper we have revisited the relation between the extended rigid body and the simple pendulum with the aim to give an exhaustive list of all different ways in which the relation takes place. We started in section \ref{RigidBody} reviewing the basics of the rigid body system in its two parameters formulation. The first parameter $e_0$ is related to the quotient $E/L^2$ where $E$ is the energy of the motion and $L^2$ is the square of the magnitude of angular momentum. The second parameter $\kappa$ codifies the values of the three moments of inertia. We work in this formulation of the rigid body because it allows us to have a good control on the different geometries of the two Casimir functions of the system \cite{delaCruz:2017qkh}.

The original construction to establish the relation between the extended rigid body and the pendulum was discussed by Holm and Marsden \cite{Marsden1} and uses the $SL(2,\mathbb{R})$ symmetry of the Euler equations to find linear combinations of the two Casimir functions and transform them to new ones, denoted in this paper as ${\cal H}$ and ${\cal N}$. For one specific class of $SL(2,\mathbb{R})$ transformations, both Casimirs have the geometry of an elliptic cylinder, this case corresponds to the extended rigid body with $ISO(2)$ Lie algebra. By a proper change of coordinates, or momentum map, and taking one of the cylinders as the cotangent space and the another one as the Hamiltonian, it is possible to obtain both the two dimensional Hamiltonian of the simple pendulum and its canonical equations of motion, even more, taking different values for the principal moments of inertia allows to get the different solutions of the simple pendulum: oscillatory, circulating and critical. Our present work revisits the construction proposed by Holm and Marsden in the two parameters formulation and, since we are working with a bi-hamiltonian system, we give the momentum maps for the two different physical situations, i.e. when ${\cal N}$ is the cotangent space and ${\cal H}$ is the Hamiltonian and the situation where we invert the role of the Casimirs, ${\cal H}$ as the cotangent space and ${\cal N}$ as the Hamiltonian. This exercise allows us to understand in a precise way what kind of movements in the rigid body are mapped to certain type of movements of the simple pendulum. Specifically, we show that movements in the rigid body, for which $e_2<e_0$, are mapped to oscillatory movements of the pendulum if  ${\cal H}$ is the Hamiltonian, but are mapped to circulating movements of the pendulum if ${\cal N}$ is the Hamiltonian instead. Something similar occurs for movements with $e_2>e_0$.

Going one step beyond and using the whole $SL(2,\mathbb{R})$ transformations of the Euler equations, in this paper we study all different cases in which there is a relation between the solutions of the extended rigid body and the ones of the simple pendulum or at least part of it.  As a result, we have found that if we keep the geometry of one of the Casimirs an elliptic cylinder and we consider the other Casimir a hyperbolic cylinder, it is also possible to get the simple pendulum. To be specific, if ${\cal H}$ represents the elliptic cylinder and ${\cal N}$ represents the hyperbolic cylinder then we have again two situations: i) If ${\cal H}$ is the cotangent space and ${\cal N}$ is the Hamiltonian we can provide a momentum map that gives origin to the whole movements of the pendulum; this case also corresponds to an extended rigid body with Lie algebra $ISO(2)$. ii) If instead ${\cal H}$ is the Hamiltonian and ${\cal N}$ is the cotangent space, then we can provide a momentum map considering one of the coordinates as purely imaginary in such a way that we get the Hamiltonian of the pendulum, although the energies of the system correspond only to the circulating movements and not the oscillatory ones; this case corresponds to the extended rigid body with Lie algebra $ISO(1,1)$. To the best of our knowledge this is a physical situation not previously discussed in the literature. In 
every case  we also give the momentum map that relates the extended rigid body to the simple pendulum with 
imaginary time.

As for future work, because the analysis performed in this paper has been established only classically it would be 
interesting to extend the construction to the quantum framework. We expect this analysis could produce interesting relations between Lame differential equation, which governs the quantum behaviour of the rigid body, and the Mathieu differential equation, which governs the quantum behaviour of the simple pendulum. A second possible direction of research is to investigate whether or not the dynamics of the pendulum could be used also to implement one-qubit quantum gates as the rigid body does \cite{VanDamme}. 

\acknowledgments 
The work of M. de la C. and N. G. is supported by the Ph.D. scholarship program of the Universidad 
Aut\'{o}noma Metropolitana. The work of R. L. is  partially supported from CONACyT Grant No. 237351 
``Implicaciones f\'{i}sicas de la estructura del espacio-tiempo".

\bibliography{mybibfile}
\end{document}